\documentclass[
 aip,
 amsmath,amssymb,
 superscriptaddress,
 preprint,
 floatfix,
]{revtex4-1}

\usepackage[usenames, dvipsnames]{color}
\usepackage{upgreek}
\usepackage{graphicx}
\usepackage{dcolumn}
\usepackage{bm}
\usepackage{hyperref}
\usepackage{appendix}
\usepackage{CJK}

\usepackage{xcolor}
\hypersetup{
    colorlinks,
    linkcolor={blue!50!black},
    citecolor={blue!50!black},
    urlcolor={blue!80!black}
}

\usepackage[T1]{fontenc}
\usepackage{mathptmx}
\usepackage{etoolbox}

\makeatletter
\def\@email#1#2{%
 \endgroup
 \patchcmd{\titleblock@produce}
  {\frontmatter@RRAPformat}
  {\frontmatter@RRAPformat{\produce@RRAP{*#1\href{mailto:#2}{#2}}}\frontmatter@RRAPformat}
  {}{}
}%
\makeatother
\begin{document}
\preprint{AIP/123-QED}

\title{Large cavitation bubbles in the tube with a conical-frustum shaped closed end during a transient process}

\author{Zhichao Wang}
\affiliation{State Key Laboratory of Hydroscience and Engineering, and Department of Energy and Power Engineering, Tsinghua University, Beijing 100084, China}

\author{Shuhong Liu}
\affiliation{State Key Laboratory of Hydroscience and Engineering, and Department of Energy and Power Engineering, Tsinghua University, Beijing 100084, China}

\author{Bo Li }
\affiliation{College of Robotics, Beijing Union University, Beijing 100020, China}
\affiliation{Key Laboratory for Thermal Science and Power Engineering of Ministry of Education, Tsinghua University, Beijing 100084, China}

\author{Zhigang Zuo*}
\email{zhigang200@tsinghua.edu.cn}
\affiliation{State Key Laboratory of Hydroscience and Engineering, and Department of Energy and Power Engineering, Tsinghua University, Beijing 100084, China}

\author{Zhao Pan*}
\email{zhao.pan@uwaterloo.ca}
\affiliation{Department of Mechanical and Mechatronics Engineering, University of Waterloo, University of Waterloo, 200 University Avenue West, Waterloo, Ontario N21, 3G1, Canada}

\date{\today}

\begin{abstract}
\textbf{Abstract:} The transient process accompanied by extreme acceleration in the conical sections of hydraulic systems (e.g., draft tube, diffuser) can induce large cavitation bubbles both at the closed ends and in the bulk liquid. The collapses of the large  cavitation bubbles can cause severe damage to the solid walls. We conduct experiments in the tubes with different conical-frustum shaped closed ends with the `tube-arrest' method and observe bubbles generated at these two locations. For the bubbles generated at the close end of the tube, we propose the onset criteria, consisting of two universal non-dimensional parameters $Ca_1$ and $Ca_2$, of large cavitation bubbles separating the water column. We investigate their dynamics including the collapse time and speed. The results indicate that the larger the conical angle, the faster the bubbles collapse. For the bubbles generated in the bulk liquid, we numerically study the collapse time, the jet characteristics and the pressure pulse at bubble collapse. We observe a much stronger jet and pressure pulse of bubbles in tubes, comparing with a bubble near an infinite plate. Our results can provide guidance in the design and safe operation of hydraulic machinery with complex geometries, considering the cavitation during the transient process.
\end{abstract}

\maketitle  

\section{Introduction}\label{sec:intro}

In engineering applications, pipes with tapered sections, such as draft tube cones behind hydraulic turbine runners, Venturi flow meters, and diffusers are widely used. 
Cavitation in these conical sections can be induced by 1) excessively high flow speed, such as the cavitation vortex rope in the draft tube \cite{Nishi}, the single cavity \cite{Mishra} and attached cavitation \cite{Barre,Haochen} in Venturi nozzles, causing pressure fluctuations and cavitation erosions; 2) extreme acceleration during the transient process (e.g., the sudden valve closing). In latter instance, cavitation bubbles can grow to sizes comparable to the pipe diameter and cause water column separation. The collapses of these bubbles may result in large scale damage to the structures, threatening the safety of the hydraulic system \cite{Bonin,Bergant2006}.

Previous studies have been carried out on the large cavitation bubbles originated next to the runner in the draft tube during the load rejection or other transient events in hydroelectric stations~\cite{Nonoshita,Pejovic2004,Pejovic2011,Zhang,He}.
The results indicate that the onset of large cavitation bubbles is related to the operating factors (e.g., the closing process of the guide vanes or runner \cite{Nonoshita}, the speed control of the runner \cite{Pejovic2004,Pejovic2011}) and the pipeline geometrical parameters (e.g., the length of the tailrace tunnel) \cite{Nonoshita}. 
Pressure surges in the hydraulic systems accompanied by the collapses of cavitation bubbles have also been studied in specific transient processes. However, the dynamics of large transient cavitation bubbles in the conical sections (e.g., bubble sizes, evolution of the bubble shape, oscillation periods, and collapse speeds) remain unclear.

In simpler geometries, i.e., straight cylindrical pipelines, cavitation bubbles generated at the closed end (e.g., the closing valve) during the transient process have been studied by both numerical and experimental methods \cite{Sharp,Simpson,Brunone,Bergant1999,Adamkowski2012,Adamkowski2015}.
It is shown that the oscillation period and size of the cavitation bubbles are determined by the initial flow velocity and the pressure differences in the system \cite{Bergant2006,Simpson,Bergant1999}. 
However, due to the destructive nature of the phenomena \cite{Bonin,Bergant2006,Jesse}, systematical experimental studies using circulating pipe system remain difficult and limited. 
In contrast, the so-called `tube-arrest' method provides an equivalent and low-cost experimental technique for the study of cavitation bubbles during the transient process \cite{Chesterman,Chen,Xu}.  In a tube-arrest setup, a tube filled with liquid is driven upwards until it impacts with a buffer, causing rapid liquid deceleration and inducing cavitation bubbles at the bottom of the tube. Based on the experiments in a tube-arrest apparatus, Xu \emph{et al.} \cite{Xu} proposed the onset criteria of large cavitation bubbles generated at the close end of the cylindrical tube. They found that the large cylindrical bubbles collapse at a finite speed, and the collapse time can be estimated by a newly established Rayleigh-type model.

In addition, due to the propagation of the rarefaction waves along the pipeline, large transient cavitation bubbles may appear not only at the quick-closing valve, but also at locations away from the valve \cite{Sharp,Bergant1999,Adamkowski2012}. The characteristics of the large cavitation bubbles generated in the bulk liquid in the middle of the tube remain unclear and deserve our further investigation.

In this paper, we systematically study the dynamics of the large cavitation bubbles in the tubes with different conical-frustum shaped closed ends using the similar tube-arrest method described in Xu \emph{et al.} \cite{Xu}. We observe cavitation bubbles generated both at the closed end and in the bulk liquid of the tubes. For the bubbles generated at the closed end of the tube,  we propose a universal model for tubes with different cone angles, to describe the onset criteria and dynamics of large cavitation bubbles. For the large bubbles generated in the bulk liquid, we observe the jetting phenomenon of collapsing bubbles, and further investigate the jet characteristics and the pressure impulse resulting from the bubble collapse. The flow focusing effects in the conical geometry accelerate the bubble collapse and the jet, potentially leading to more damage to the structures.

\section{Experimental Setup}

\begin{figure*}[t]
\centering
\includegraphics[width=\linewidth]{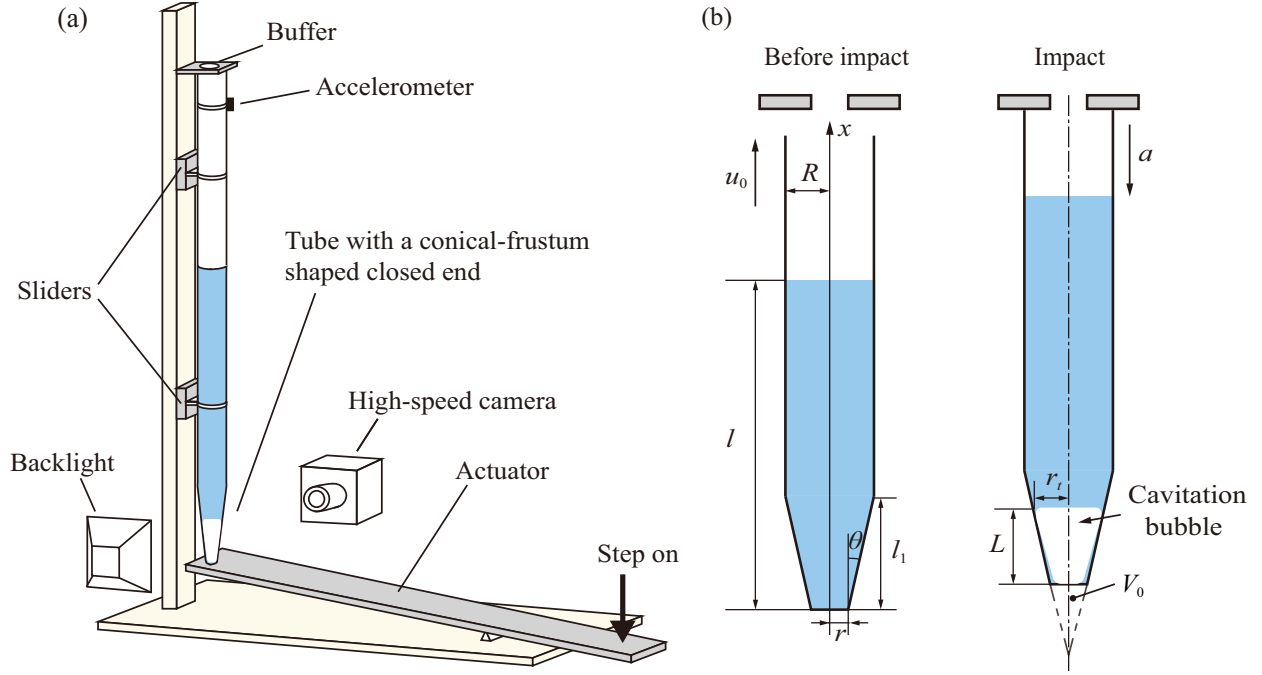}
\caption{Experimental setup and notations. (a) Schematic of the experimental apparatus. (b) Cavitation bubble generation  by tube-arrest principle and major parameters (dimensions not to scale). The arrows beside $u_0$ and $a$ indicate their directions.}
\label{setup}
\end{figure*}

In the current research, the modified tube-arrest setup, adapted from previous work by Xu \emph{et al.} (see also Fig.~\ref{setup}) \cite{Xu}, was used to generate and study the cavitation behaviour.
An acrylic tube with a conical-frustum shaped closed end filled with deionized water is driven upwards by stepping on the actuator until the tube hits the bottom of the stopper, where buffer material is attached. 
After the impact, the tube is arrested by the buffer while the liquid column continues moving upwards due to inertia. 
Cavitation bubbles may thus be induced when the tension in the liquid exceeds the tensile strength.

A high-speed camera (Phantom V711, Vision Research, USA) was used to record the cavitation bubbles with a frame rate of 20,000~fps and an exposure time of 10~$\upmu$s. 
An LED light with a diffuser was used for illumination. 
The tube acceleration upon impact $a$ was measured by an accelerometer (357B03, PCB, USA) attached on the top of the tube sampling at 102,400 Hz with the uncertainty of 2$\%$. 
The accelerometer and the high-speed camera were triggered synchronously by a delay generator (9524, Quantum composers, USA). 
The displacement of the tube and the bubble length $L$ were measured from the calibrated high speed images. 
The tube impact velocity $u_0$ was adjusted by how hard one steps on the actuator, and is calculated by linear regression of the displacement of the tube $\sim 3$~ms before impact, with an uncertainty of 0.03~m/s. 
In this setup, the tube acceleration $a$ upon impact was decoupled from the tube impact velocity $u_0$, by alternating the materials of the buffer (e.g., rubber, plastic board, marble, and ceramic tile) and thicknesses. 
The resulting $u_0$ and $a$ lay in the range of 0.29~--~2.67~m/s and 88~--~2048~$\mathrm{m/s^2}$, respectively, in our experiments.

In order to investigate the influence of the tube geometry on the cavitation bubble dynamics, we built six tubes with conical-frustum shaped bottoms. 
The  tubes have different half cone angles $\theta$ varying from $\theta = 0^{\circ}$ to $\theta = 45^{\circ}$, and the  parameters are listed in Table~\ref{table1}.  Note, $\theta = 0^{\circ}$ corresponds to a straight cylindrical tube. The total length and wall thickness of the tubes are 1.0~m and 3.0~mm, respectively. Considering the length ratio of the conical to cylindrical section of a draft tube, and to avoid excessive tube diameter, $l_1$ is designed to be relatively small compared to the total tube length. In the experiments, the heights of water column $l$ were nominally set as 200~mm, 300~mm, 400~mm, 550~mm, and 600~mm. 
The uncertainties of the tube diameter and liquid column length are 0.1~mm and 1.0~mm, respectively.
The measurement error of the bubble length obtained from the images is less than 20 pixels, and 10 pixels in most cases (approximately 1.7~mm in physical dimension). 

It should be noted that the effect of potential nucleation sites in the bulk liquid is significant. 
We machined and polished the conical-frustum section of each tube as a whole, so that the inner surface at the tube bottom is much smoother than the ones used in previous research of Xu \emph{et al.}~\cite{Xu}. Thus, aside from bubbles generated on the tube bottom, we expect to observe the bubbles originated in the bulk liquid due to the absence of surface nuclei.

\begin{table}[htbp]
\caption{\label{table1}Parameters of the six tubes with a conical-frustum shaped closed end.}
\begin{ruledtabular}
\begin{tabular}{ccccc}
Tube& $\theta~({^\circ})$&$r$~(mm)&$R$~(mm)&$l_1$~(mm)\\
\hline
1&0&9.5&9.5&-\\
2&6&5.0&17.0&114.2\\
3&15&5.0&17.0&44.8\\
4&30&5.0&17.0&20.8\\
5&45&5.0&17.0&12.0\\
6&45&2.0&24.5&22.5\\
\end{tabular}
\end{ruledtabular}
\end{table}

\section{Experimental observations}
\label{sec:Observations}

\begin{figure*}[htb]
\centering
\includegraphics[width=\linewidth]{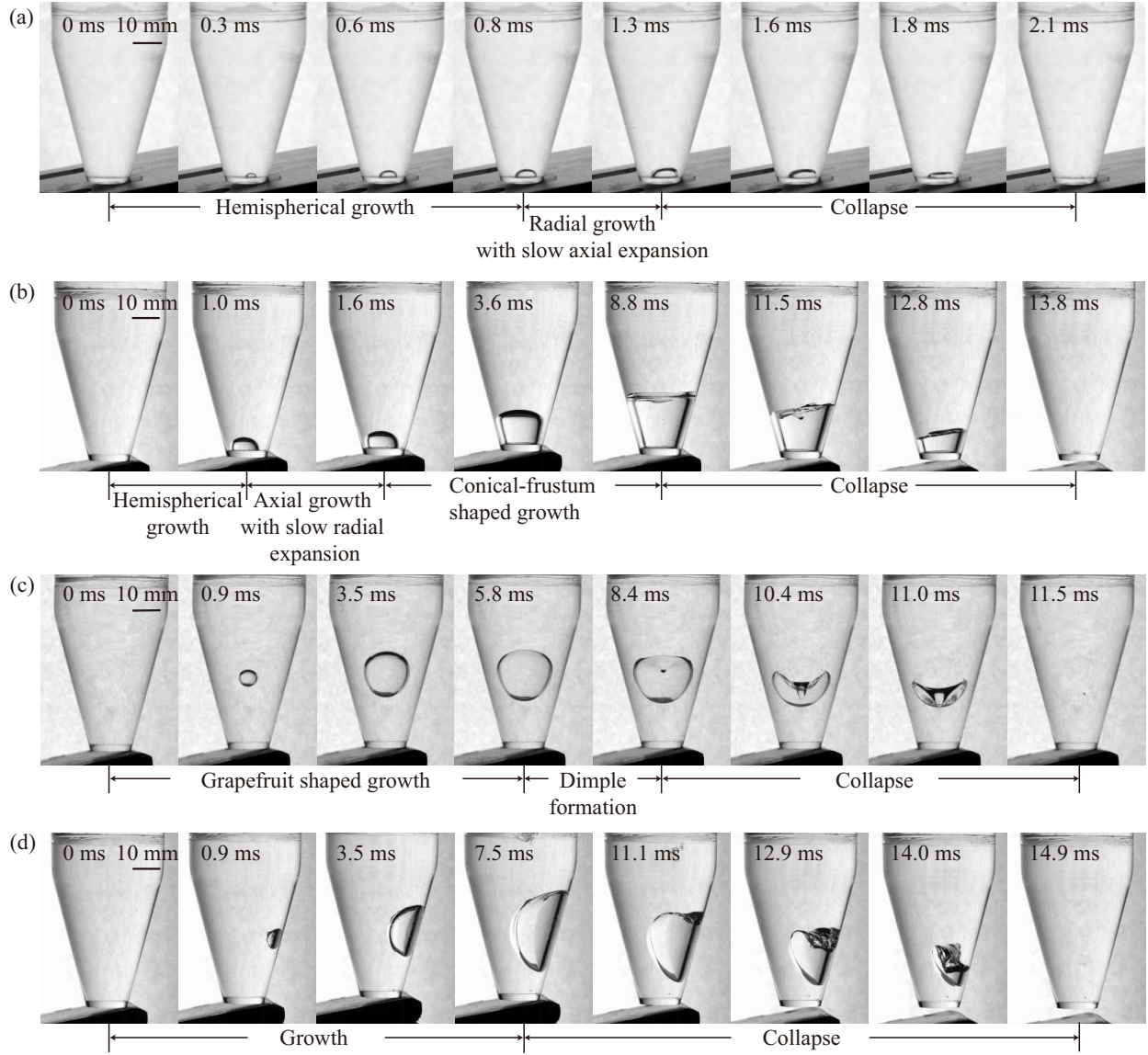}
\caption{High-speed images of typical cases of cavitation bubble(s) formed at the bottom of the tube and in the bulk liquid. Here, $t = 0$~ms indicates the moment of tube impacting upon the buffer. The half cone angle $\theta=15^{\circ}$ for all the four cases. (a) Small bubble generated on the tube bottom (multimedia view). $u_0=0.71~\mathrm{m/s}$ and $a=1158.95~\mathrm{m/s^2}$. Height of the water column $l=200$~mm. (b) Large bubble generated on the tube bottom (multimedia view). $u_0=2.11~\mathrm{m/s}$, $a=681.10~\mathrm{m/s^2}$, and $l=400$~mm. (c) Bubble originated in the bulk liquid in the middle of the conical-frustum section (multimedia view). $u_0=1.21~\mathrm{m/s}$, $a=585.50~\mathrm{m/s^2}$, and $l=550$~mm. (d) Bubble originated on or near the side wall (multimedia view). $u_0=2.31~\mathrm{m/s}$, $a=735.00~\mathrm{m/s^2}$, and $l=400$~mm.}
\label{sequence}
\end{figure*}

In this section, we will present the experimental results and discuss the onset of the large cavitation bubbles. 
Figure~\ref{sequence} shows typical high-speed images of cavitation bubbles generated in the conical-frustum shaped tube. 
In all the cases, the onset of the bubble occurs immediately after the tube is arrested (at time $t = 0 $~ms). Figures~\ref{sequence}(a) and (b) present images of typical evolution of bubbles generated at the tube bottom. 
In Fig.~\ref{sequence}(a) (multimedia view), the bubble firstly undergoes a hemispherical growth (0~--~0.8~ms) until  the edge of the bubble contacts the side wall of the tube (at $\sim0.8$~ms).  Then the bubble grows mostly in the radial direction along the tube with a slow expansion in the axial direction, to its maximum size (0.8~--~1.3~ms). The bubble's first collapse occurs afterwards (1.3~--~2.1~ms). In this case, the bubble is small, and the water column separation does not occur during the entire growth and collapse phases of the bubble.

A water column separation case is presented in  Fig.~\ref{sequence}(b) (multimedia view), where the bubble is much larger than the one showed in Fig.~\ref{sequence}(a).
The bubble  grows in the shape of a hemisphere initially (0~--~1.0 ms). 
When the bubble approaches the side wall of the tube, the bubble experiences a slow expansion in the radial direction, until the bubble fills up the bottom of the tube (1.0~--~1.6 ms). 
Afterwards, it develops into a conical-frustum shape, featuring a full water column separation, and continues to grow to its maximum size due to the geometrical confinement (1.6~--~8.8 ms). 
The bubble then undergoes its first collapse (8.8~--~13.8~ms).

In order to identify the conditions for the occurrence of water column separation, we hereby define that the bubble generated at the bottom of the tube with maximum length larger than the diameter of the bottom ($L_{\max}>2r$, as shown in Fig.~\ref{sequence}(b)) is a `large cavitation bubble'. This is in analogy to the categorization proposed by Xu \emph{et al.}~\cite{Xu}. 
In the experiments, we observe that water column separation is almost guaranteed to happen if such a large cavitation bubble appears.

We also include two examples where bubbles are in the bulk liquid (Figs.~\ref{sequence}(c) and (d)).
When the bubble is originated in the middle of the conical section of the tube as shown in Fig.~\ref{sequence}(c) (multimedia view), it develops into a tapered shape as it grows, with a slightly narrowed bottom and a flattened top (0~--~5.8~ms).
During its collapse, the top of the bubble deform and forms a dimple at the center (5.8~--~8.4~ms). 
The dimple cusp then develops into a strong downward-shooting jet (8.4~--~11.5~ms). It should be noted that the jetting phenomenon was rarely seen when the bubble was generated at the tube bottom in our experiments. Figure~\ref{sequence}(d) presents the evolution of a bubble originated on or near the side wall surface of the tube (multimedia view).
We observe that it grows along the side wall to its maximum size (0~--~7.5~ms), and collapses with complex deformation and jetting (7.5~--~14.9~ms), due to the effects of nearby boundaries.

It is seen that the bubble generated in the bulk liquid in the middle of the tube in Fig.~\ref{sequence}(c), with a maximum volume of about $3.59\times10^{-6}~\mathrm{m^3}$, does not induce water column separation. 
However, it already exceeds the critical volume for large bubbles $1.28\times10^{-6}~\mathrm{m^3}$ defined above (i.e., $L_{\max}>2r$) if it appeared at the tube bottom, given the half cone angle $\theta=15^{\circ}$ and the radius of the tube bottom $r=5.0~\mathrm{mm}$. 
In fact, in our experiments ($1.03~\mathrm{m/s}<u_0<2.68~\mathrm{m/s}$, and $200~\mathrm{mm}<l<600~\mathrm{mm}$), we rarely observe the large cavitation bubbles generated in the bulk liquid separating the liquid column, because the bubbles in the middle of the tube are more difficult to fill the whole space under the same conditions.

Therefore, the discussion on the liquid column separation in the following sections, we will focus on the bubbles generated at the bottom of the tube. 
Since it is well known that the cavitation bubble collapse near a solid boundary can induce high pressure pulses and cause potential damage to the wall \cite{Benjamin,Philipp}, some discussions on the mechanism and characteristics of large bubbles generated in the bulk liquid (as exemplified in Fig.~\ref{sequence}(c)) are useful, and are provided in Section~\ref{sec:middlebubble}.

\section{Large bubbles at tube bottom}
\label{sec:bottombubble}

Now we investigate the onset criteria of large cavitation bubbles generated at the bottom. Here, ``large'' is quantified by the bubble length, which is long enough to yield liquid column separation in a tube (adapted from \cite{Xu}). The onset criteria consist of two simultaneous conditions: i) cavitation bubble occurs, and ii) the cavitation bubble can grow to a large size (i.e., $L_{\max}>2r$). We hereby show that these two requirements can be described by two independent non-dimensional parameters.

The first parameter shall dictate the criteria of cavitation onset. Recall that Pan \emph{et al.} \cite{Pan} proposed a cavitation number describing the onset criteria in a straight cylindrical tube in transient scenarios, where the influence of liquid acceleration on pressure variation is much greater than that of the flow velocity. 
The cavitation number was developed based on linear and one-dimensional simplifications, which may fail for complex geometries. Thus, we attempt to derive a universal expression for tubes with varying cross-sectional areas.

First, we discuss the influence of the viscosity on liquid motion. As analyzed in Onuki \emph{et al.} \cite{Onuki} and Xu \emph{et al.} \cite{Xu}, the thickness of the boundary layer $\delta$ at the side wall of the tube can be estimated as $\delta=\sqrt{\nu t}$, where $\nu$ is the kinematic viscosity of the fluid and $t\sim 2R/u_0$. Given that $\nu\sim O(10^{-6})$~$\mathrm{m^2/s}$, $u_0\sim O(1)$ m/s, and $R\sim O(10^{-2})$~m, $\delta/R\sim O(10^{-2})$. The thickness of the boundary layer is negligible compared with the tube radius. Moreover, for most cases in our experiments, the interface between the liquid and the bubble generated at the tube bottom remains flat but not quadratic during the collapse, as illustrated in Fig.~\ref{sequence}(b). This indicates a uniform velocity distribution in the liquid along the radial direction. Thus we can neglect the viscosity and apply a quasi-one-dimensional model in the following discussion.

Ignoring the liquid viscosity, the quasi-one-dimensional Euler equation of the liquid along the longitudinal direction ($x$) of the tube is
\begin{equation}
    \rho_l A(x)\frac{\partial u}{\partial t}=-\frac{\partial (pA(x))}{\partial x},
\label{N-S}
\end{equation}
where $\rho_l$ is the density of the liquid, $A(x)$ is the cross-sectional area of the tube at different $x$, and $\partial u/\partial t$ is the liquid acceleration upon impact. Integrating Eq.~\eqref{N-S} along $x$-direction from the free surface to the bottom of the liquid column gives
\begin{equation}
    p_\infty A-p_bA_b=\rho_l a V_l,
    \label{pr-pb}
\end{equation}
where $p_\infty$ and $p_b$ are respectively the atmospheric pressure and the pressure at the bottom of the column; $A=\pi R^2$ and $A_b=\pi r^2$ are the area of free surface and the bottom of the tube, respectively, and $V_l$ is the volume of the liquid column. 
Note that the liquid acceleration is approximated by the tube acceleration upon impact $a$ in the equation. 
The justification of this approximation can be referred to Xu \emph{et al.} \cite{Xu}. 
Although the pure water can withstand tension, water in our ``dirty'' experimental system cannot, so the cavitation tends to occur when $p_b<p_v$ ($p_v$ is the saturated vapor pressure). 
We then define a non-dimensional number
\begin{align*}
    Ca_1=\frac{p_\infty A-p_vA_b}{\rho_l aV_l},
\end{align*}
and $Ca_1<1$ suggests that cavitation is likely to occur. 
Since $p_v \ll p_\infty$ and $A_b<A$, $Ca_1$ can be simplified as
\begin{equation}
    Ca_1=\frac{p_\infty}{\rho_l a\frac{V_l}{A}}.
\label{Ca1-cone}
\end{equation}
For a straight cylindrical tube, Eq.~\eqref{Ca1-cone} readily reduces to $Ca_1= \frac{p_\infty}{\rho_l a l},$ where $l={V_l}/{A}$ is the length of the liquid column, and is identical to the result proposed in Pan \emph{et al.} \cite{Pan}. 
  
\begin{figure}[t]
\centering
\includegraphics[width=3.5in]{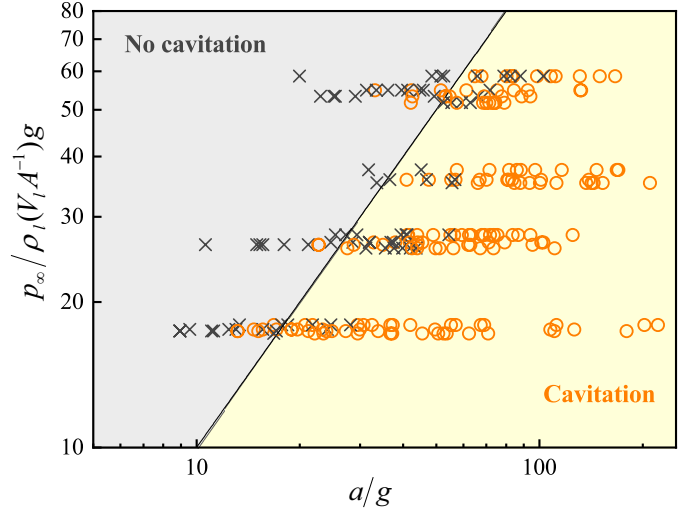}
\caption{Phase diagram for the cavitation onset by acceleration for various $V_l/A$. Orange open circles denote cavitation detection and grey symbols denote absence of cavitation detection. The black line represents theoretical prediction separating  cavity formation (lower right) and none (upper left) based on $Ca_1 = 1$ in Eq.~\eqref{Ca1-cone}}.
\label{Ca1}
\end{figure}

The experimental validation of Eq.~\eqref{Ca1-cone} with different $V_l/A$, ranging from 176~mm to 600~mm, is presented in Fig.~\ref{Ca1}. More specifically, we set the length of the liquid column $l=$ 200~mm, 300~mm, 400~mm, and 600~mm in tubes with different half cone angles ($\theta$), and suppose $p_\infty = 101,325$~Pa and $\rho_l = 1,000$~$\mathrm{kg/m^2}$ at room temperature. 
In the current research, we detect the cavitation occurrence by observing bubbles equal to or larger than 1~pixel ($\sim0.17$~mm) in the high-speed images. 
As shown in Fig.~\ref{Ca1}, the orange symbols indicating cavitation onset are mainly distributed in the lower right region ($Ca_1 < 1$), and grey symbols indicating no cavitation cases are mostly clustered in the upper left region ($Ca_1 > 1$). 
It is seen that the onset of transient cavitation in the tube can be well predicted by $Ca_1$ with different conical-frustum shaped closed ends. 
According to Eq.~\eqref{Ca1-cone}, we know that $Ca_1$ is proportional to the sectional area $A$ and inversely proportional to the volume of the liquid column $V_l$ as well as the acceleration $a$.  
Thus, for a transient process encountered in engineering scenarios (e.g., rapid closing of valves), to avoid cavitation onset, it is favorable to have a short and thick tube, and a moderate valve closing process for mild liquid acceleration.

The second non-dimensional parameter is derived by evaluating the volume of the cavitation bubbles generated in the transient process. 
In the experiments, the kinetic energy of the liquid column $E_k = 
\frac{1}{2}\rho_l V_l{u_0}^2$ before the tube collides with the buffer, where $V_l$ is the volume of the liquid column. The liquid velocity can be approximated by the tube impact velocity \cite{Xu}. 
After  collision, cavitation bubble(s) begins to grow near the bottom of the tube. The velocity of liquid then approaches zero when the bubble(s) reaches the maximum volume $V_{\max}$.
The corresponding potential energy of the bubble(s) can be evaluated as $E_p = (p_\infty-p_v) V_{\rm max}=p_\infty V_{\max}$, given that $p_\infty \gg p_v$.
Neglecting energy losses during this process (e.g., due to liquid viscosity) and  equating the initial liquid kinematic energy and the potential energy stored in the cavitation bubble, the normalized maximum bubble(s) volume can be evaluated as
\begin{equation}
   \frac{V_{\rm max}}{V_l}=\frac{0.5\rho_l {u_0}^2}{p_\infty} = \frac{1}{Ca_0},
\label{Eq:Vbmax}
\end{equation}
where $Ca_0 = {p_\infty}/0.5 \rho_l u_0^2$ is the classical cavitation number, whose velocity scale refers to the steady motion of the liquid before the collision of the tube on buffer. 
For a straight cylindrical tube,  Eq.~\eqref{Eq:Vbmax} can be simplified as the same expression in Xu \emph{et al.} \cite{Xu}
\begin{align*}
   \frac{L_{\rm max}}{l}=\frac{0.5\rho_l {u_0}^2}{p_\infty},
\end{align*}
where $l$ is the length of the water column. 
Note, although the transient cavitation onset is caused by a large liquid acceleration~$a$, the maximum volume of the bubbles is irrelevant to $a$.
Instead, it is mainly determined by the initial velocity of the flow and the volume of the liquid column.

\begin{figure}[t]
\centering
\includegraphics[width=3.5in]{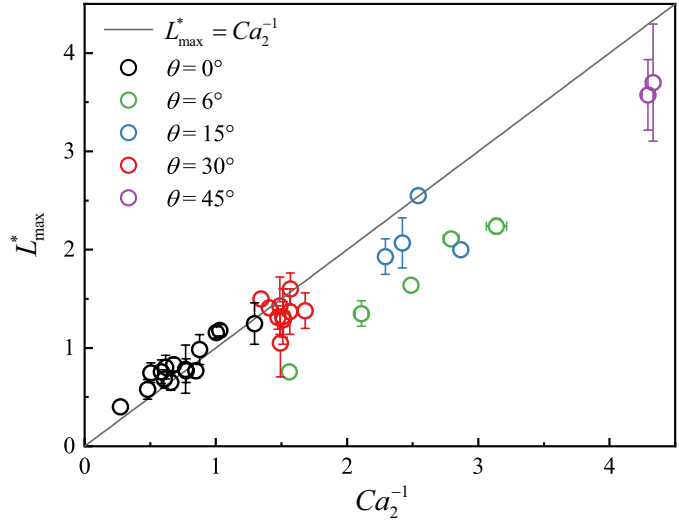}
\caption{Experimental validation of Eqs.~\eqref{Ca2-cone} and \eqref{Ca2-round}, i.e., $L_{\max}^*=Ca_2^{-1}$.}
\label{Ca2}
\end{figure}

Now we investigate the maximum bubble length $L_{\rm max}$ of the conical-frustum shaped bubble on the tube bottom. 
The maximum bubble volume $V_{\max}=\pi L_{\max}(r^2+rr_t+{r_t}^2)/3$, where $r_t=r+L_{\max}{\rm tan}\theta$ is the radius of the top of the bubble.  
Rearranging the espression of $V_{\max}$, a new non-dimensional number can be defined as
\begin{equation}
   Ca_2=L_{\rm max}^{*-1}
   =\frac{2{\tan}\theta}{\sqrt[3]{\frac{1}{Ca_0}\frac{V_l}{V_0}+1}-1}
\label{Ca2-cone}
\end{equation}
for $0^{\circ}< \theta <90^{\circ}$, where $L_{\rm max}^*=L_{\rm max}/2r$, $V_0 = \pi r^3/(3{\tan}\theta)$ is the volume of the virtual conical end indicated in Fig.~\ref{setup}(b), and $V_l$ is the volume of the water column. This parameter is determined by the classical cavitation number $Ca_0$, the half cone angle $\theta$, the radius of the bottom of the tube $r$, and the volume of the water  $V_l$.
Recall that the water column separation is almost guaranteed to occur when $L_{\rm max}>2r$,  so $Ca_2=L_{\rm max}^{*-1}<1$ is the onset criterion of a large bubble originated at the bottom of the tube. In the limit of $\theta \to0^{\circ}$, i.e., for a straight cylindrical tube,  Eq.~\eqref{Ca2-cone} approaches the same expression in Xu \emph{et al.} \cite{Xu} as below
\begin{equation}
   Ca_2=L_{\rm max}^{*-1}=\frac{2r}{l}\frac{p_\infty}{0.5\rho_l {u_0}^2}=\frac{2r}{l} Ca_0,
\label{Ca2-round}
\end{equation}

The dependence of  $L_{\rm max}^*$ on the cavitation number $Ca_2$ (i.e., Eq.~\eqref{Ca2-cone} and Eq.~\eqref{Ca2-round}) is validated in Fig.~\ref{Ca2} against the experimental results from tubes with different $\theta$. We have to point out that this bubble length relation holds only for conical-frustum shaped bubbles generated at the bottom of the tube (as in Figs.~\ref{sequence}(a) and (b)), and is not suitable for bubbles in the bulk liquid (as in Figs.~\ref{sequence}(c) and (d)).

\begin{figure}[t]
\centering
\includegraphics[width=3.5in]{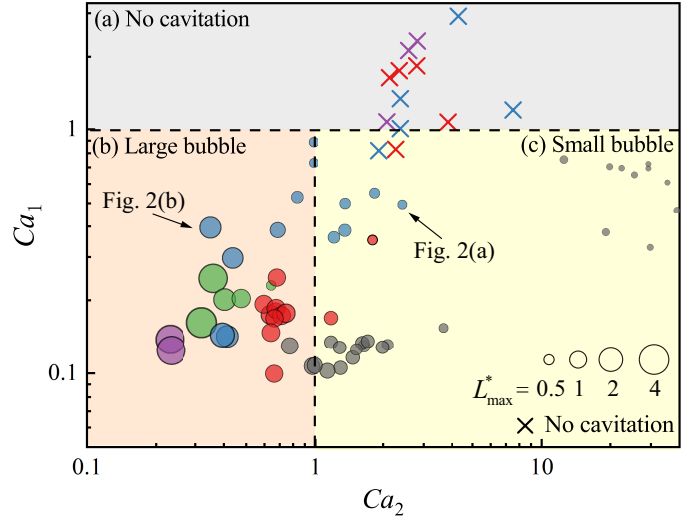}
\caption{Non-dimensional length $L_{\max}^*$ of  conical-frustum shaped cavitation bubbles originated at the bottom of the tube  on $Ca_1$ \rm{vs} $Ca_2$ phase diagram. The sizes of the symbols indicate the values of $L_{\max}^*$. Grey, green, blue, red, and purple symbols represent cases of $\theta=0^{\circ}$, $6^{\circ}$, $15^{\circ}$, $30^{\circ}$, and $45^{\circ}$ respectively. Black arrows indicate the cases in Figs.~\ref{sequence}(a) and (b).}
\label{Ca2phase}
\end{figure}

The criteria discussed above for tubes with different $\theta$ are validated against the experimental data on the $Ca_1$ \rm{vs} $Ca_2$ diagram, as shown in Fig.~\ref{Ca2phase}.
It is seen that the diagram is divided into three regimes by $Ca_1 =1$ and $Ca_2 =1$. 
First, we observe that almost all the cases with no cavitation are distributed in the $Ca_1>1$ region (Regime (a)). Then the $Ca_1<1$ region is further divided into two sub-domains (b) and (c). 
In Regime (b), i.e., $Ca_1<1$ and $Ca_2<1$, we observe the occurrence of large cavitation bubbles ($L_{\max}^*>1$), and the water column separation phenomena are almost guaranteed to happen. Most events with small cavitation bubbles ($L_{\max}^*<1$) are located in regime (c), where $Ca_1<1$ and $Ca_2>1$. 

Based on the above analysis, the onset of large cavitation bubbles originated at the bottom of the tube requires both $Ca_1 < 1$ (onset threshold for cavitation, small or large) and $Ca_2 < 1$ (threshold for large bubbles) simultaneously. In engineering contexts, this indicates that in order for the transient cavitation to occur in a tube in a hydraulic system, the transient process needs a large enough liquid acceleration (a small $Ca_1$), e.g., caused by a fast closing valve or guide vanes. However, once the cavitation bubbles are generated, whether they can grow to large sizes that induce water column separations, is no longer influenced by the severity of the transient process, but determined by the steady motion of the liquid before the transient process (a small $Ca_2$). 
Meanwhile, the specific geometry of the tube with a conical-frustum shaped closed end affects the form of both $Ca_1$ and $Ca_2$: $Ca_1$ increases with the radius at the top of the conical frustum ($R$), and $Ca_2$ is closely related to the half cone angle $\theta$ and the bottom radius $r$ of the conical frustum.

Now we discuss the oscillation period and the collapse speed of the cavitation bubbles that are essential for the estimation of the magnitude of the pressure pulses in the transient processes. 
In our experiments, it is commonly seen that for the large cavitation bubble originated at the bottom of the tube, it remains almost conical-frustum shaped  with a flat vapor-water interface  during its collapse, which is helpful in theoretical analysis.

Ignoring the mass diffusion between the gas and liquid phases, and the influence of the boundary layer flow at the side wall of the tube, the continuity equation of the incompressible liquid reads 
\begin{equation}
    \pi{R}^2\frac{{\rm d}l}{{\rm d}t}= \pi{r_t}^2\frac{{\rm d}L}{{\rm d}t},
\label{continuty}
\end{equation}
where $R$ and $r_t=r+L \cdot {\tan}\theta$ are respectively the inner radius of the straight section of the tube and the  radius of the vapor-liquid surface (see Fig.~\ref{setup}(b)), and ${{\rm d}l}/{{\rm d}t}$ and ${{\rm d}L}/{{\rm d}t}$ are respectively the velocities of the free surface and the vapor-liquid interface. 
Invoking the axial component of the momentum equation (Eq.~\eqref{pr-pb}), where the liquid viscosity is ignored, we derive a Rayleigh-type equation for the dynamics of the collapse of the conical-frustum shaped bubble:
\begin{equation}
    \frac{{\rm d^2}l}{{\rm d}t^2}=-\frac{\pi{R}^2p_\infty}{\rho_l V_l},
\label{momentum}
\end{equation}
where $V_l$ is the volume of the liquid column, and ${{\rm d^2}l}/{{\rm d}t^2}$ is the acceleration of the free surface.

Applying initial conditions $t=0$ and  ${{\rm d}l}/{{\rm d}t|_{t=0}=0}$, integrating Eq.~\eqref{momentum} gives $\frac{{\rm d}l}{{\rm d}t}=-\frac{\pi{R}^2p_\infty}{\rho_l V_l}t$, which can then be combined with Eq.~\eqref{continuty}, the bubble length $L$ during the collapse stage can be described as 
\begin{equation}
   \pi(r+L \cdot \tan \theta)^2{\rm d}L=-\frac{\pi^2{R}^4p_\infty}{\rho_l V_l}t{\rm d}t.
\label{Eq7+8}
\end{equation}
In this equation $ \pi(r+L\cdot \tan \theta)^2{\rm d}L$ is the differential form of the cavitation bubble volume $V_b$. Given initial conditions that $L|_{t=0}=L_{\max}$ and $V_b|_{t=0}=V_{\max}$, integration of Eq.~\eqref{Eq7+8} leads to
\begin{equation}
   V_b=-\frac{A^2p_\infty}{2\rho_l V_l}t^2+V_{\max},
\label{Vb-t}
\end{equation}
where $A=\pi R^2$ is the cross-sectional area of the straight section of the tube, and the maximum bubble volume $V_{\max}=0.5\rho u_0^2V_l/p_\infty$ according to Eq.~\eqref{Eq:Vbmax}.

Here we define the duration from the cavitation bubble at its maximum size to the first collapse as its collapse time $T_c$. 
According to Eq.~\eqref{Vb-t},   
\begin{equation}
   T_c=\frac{\rho_l u_0}{p_\infty}\frac{V_l}{A} = \sqrt{2} \frac{\sqrt{V_l V_{\max}}}{A} \sqrt{\frac{\rho_l}{p_{\infty}}}.
\label{Tc-cone}
\end{equation}
In the case of a straight cylindrical tube ($\theta \to 0^{\circ}$ and $V_l=Al$), Eq.~\eqref{Tc-cone} reduces to
\begin{align*}
   T_c=\frac{\rho_l u_0}{p_\infty}l = \sqrt{2} \sqrt{l L_{\max}} \sqrt{\frac{\rho_l}{p_{\infty}}},
\end{align*}
which is consistent with the expression in Xu \emph{et al.}~\cite{Xu}. 

\begin{figure}[t]
\centering
\includegraphics[width=3.5in]{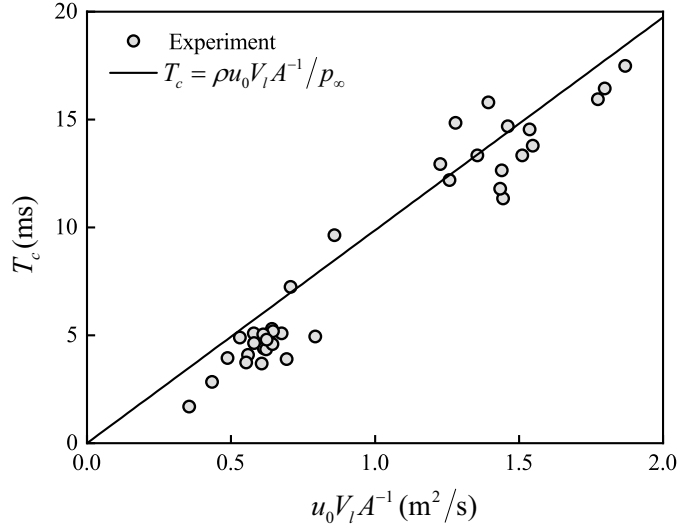}
\caption{Experimental validation of Eq.~\eqref{Tc-cone}, $T_c=\frac{\rho_l u_0}{p_\infty}\frac{V_l}{A}$. Error bars are smaller than the size of data points.}
\label{Tc}
\end{figure}

The experimental validation of Eq.~\eqref{Tc-cone} is presented in Fig.~\ref{Tc}. It is seen that given the liquid density $\rho_l$ and the atmospheric pressure $p_{\infty}$, the collapse time $T_c$ of a conical-frustum shaped cavitation bubble generated on the tube bottom is determined by the steady flow velocity before the transient process $u_0$, the volume of the liquid column $V_l$, and the cross-sectional area $A$ of the straight section of the tube. 
This result is particularly useful in engineering, since the parameters needed for estimating $T_c$ are either geometrical factors of the tube, or bulk flow velocity that can be derived by the volume flow rate going through the tube.
In comparison with a classic Rayleigh bubble in an infinite liquid region, whose collapse time is only determined by the maximum bubble size given $\rho_l$ and $p_{\infty}$, the conical-frustum shaped bubble has a $T_c$ related to the geometric average of $V_l$ and $V_{\max}$, and the diameter of the tube.

Now we focus on the collapse speed of the cavitation bubble. As analyzed in Xu \emph{et al.}~\cite{Xu}, the water-hammer pressure generated by the first collapse of the bubble can be estimated as $\rho cu_c$, where $c$ is the speed of the sound in water, and $u_c$ is the dimensional collapse speed of the cavitation bubble. According to Eq.~\eqref{Eq7+8}, we can readily deduce that $u_c=\frac{{\rm d}L}{{\rm d}t}=\frac{R^2}{r^2}u_0$ and estimate the magnitude of the pressure impulse.

To gain further insight into the effect of different tube geometries on the collapse speed, non-dimensional analysis is required.
From Eq.~\eqref{Vb-t}, noting that $V_{b}=\pi L(r^2+rr_t+{r_t}^2)/3$ where $r_t=r+L{\rm tan}\theta$, we can obtain the non-dimensional bubble length $L^*=L/L_{\max}$ 
\begin{equation}
\begin{aligned}
   L^*&
   =\frac{\sqrt[3]{-Ct^{*2}+C+1}-1}{\sqrt[3]{C+1}-1},
\label{L*-t*-cone}
\end{aligned}
\end{equation}
where $t^*=t/T_c$, and $C=(2\tan \theta \cdot  Ca_2^{-1}+1)^3-1$ based on Eq.~\eqref{Ca2-cone}. This relation between $L^*$ and $t^*$ indicates that the collapsing dynamics of the conical-frustum shaped cavitation bubble is determined by $\tan \theta \cdot Ca_2^{-1}$. For a straight cynlindrical tube with $\theta \to0^{\circ}$, the limit of Eq.~\eqref{L*-t*-cone} gives the same expression in Xu \emph{et al.} \cite{Xu} as below 
\begin{align*}
   L^*=-t^{*2}+1.
\end{align*} 

\begin{figure}[t]
\centering
\includegraphics[width=3.5in]{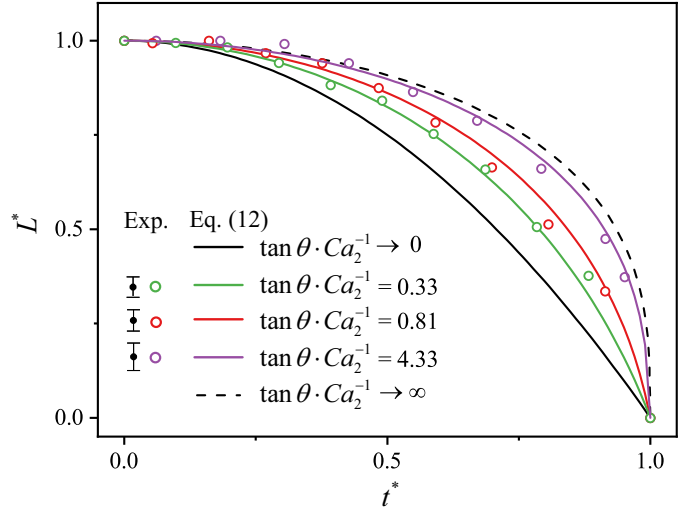}
\caption{Non-dimensional cavitation bubble length $L^*$ verses non-dimensional time $t^*$ with different $\tan \theta \cdot Ca_2^{-1}$. The error bar for each case is shown to the left of the corresponding
legend marks. Movies for $\tan \theta \cdot Ca_2^{-1}=$~0.33 (multimedia view),~0.81 (multimedia view),~4.33 (multimedia view) are available.}
\label{Dynamics}
\end{figure}

We validate Eq.~\eqref{L*-t*-cone} against experimental results in Fig.~\ref{Dynamics}. It is seen that an increasing $\tan \theta \cdot Ca_2^{-1}$ results in a larger bubble collapse speed, as theoretically predicted by Eq.~\eqref{L*-t*-cone}.
Two extreme cases are also presented with theoretical results. The case with a straight cylindrical tube ($\theta = 0^{\circ}$, hence $\tan \theta \cdot Ca_2^{-1} \to 0$) is denoted by the black solid curve in Fig.~\ref{Dynamics} and is consistent with the validated collapse dynamics in Xu \emph{et al.} \cite{Xu}.
In the limit of $\tan \theta \cdot Ca_2^{-1} \to \infty$, Eq.~\eqref{L*-t*-cone} gives
\begin{align*}
   L^*=\sqrt[3]{-t^{*2}+1},
\end{align*}
which corresponds to the black dashed curve.
$\tan \theta \cdot Ca_2^{-1} \to \infty$ represents either a hemispherical bubble on an infinite plate in a liquid ($\theta \to90^{\circ}$), or a conical bubble ($r \to0$). In both scenarios, the collapse speed of the bubble approaches infinity.

\begin{figure}[t]
\centering
\includegraphics[width=4in]{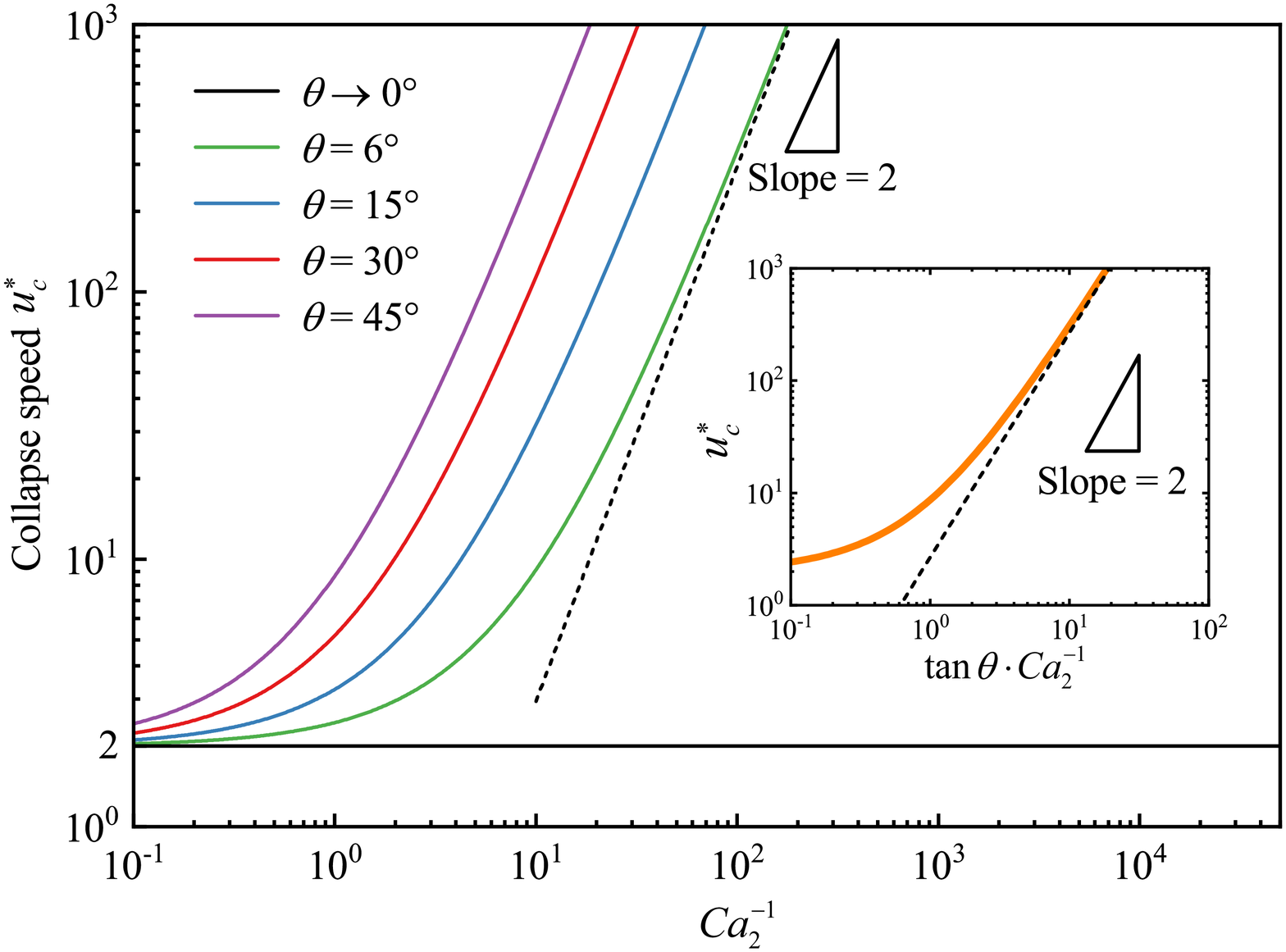}
\caption{Non-dimensional bubble collapse speed $u_c^*$ \rm versus $Ca_2^{-1}$ and $\tan \theta \cdot Ca_2^{-1}$.}
\label{Fig:uc}
\end{figure}

From Eq.~\eqref{L*-t*-cone} we can calculate the non-dimensional bubble collapse speed $u_c^* = \frac{{\rm d} L^* }{{\rm d} t^*} | _{t^*=1}$ for $\theta \in [0^{\circ}, 90^{\circ})$
\begin{align*}
   u_c^*=\frac{\frac{2}{3}C}{\sqrt[3]{C+1}-1},
\end{align*}
which is plotted in Fig.~\ref{Fig:uc} for several typical values of $\theta$ with various colors.

For a  straight cylindrical tube ($\theta \to 0^{\circ}$),  $u_c^* = 2$ and is irrelevant to $Ca_2$. When $\theta >0^{\circ}$, $u_c^*$ increases slowly with $Ca_2^{-1}$ at moderate $Ca_2 ^{-1}$ values.  With a sufficiently large  $Ca_2 ^{-1}$, $u_c^*$ grows fast, and the $u_c^*$ versus $Ca_2^{-1}$ curves approach asymptotes with the same slope of 2 for different $\theta$ (see the colored lines in Fig.~\ref{Fig:uc}), and the larger $\theta $ leads to the earlier fast growth of $u_c^*$.

Scaling $Ca_2^{-1}$ by $\tan \theta$, we find that colored lines for different $\theta$ collapse to a universal one (see the thick orange curve in the sub-graph in Fig.~\ref{Fig:uc}).
This indicates that the non-dimensional bubble collapse speed $u_c^*$ solely depends on  $\tan \theta \cdot Ca_2^{-1}$, and $u_c^*$ is proportional to $(\tan \theta \cdot Ca_2^{-1})^2$ at large values of $\tan \theta \cdot Ca_2^{-1}$. 
In engineering applications, considering that the pressure pulse at the bubble collapse increases with its collapse speed, we can thus conclude that a large $Ca_2$ can both reduce the risk of water column separation, and the magnitude of pressure pulse. 
According to Eq.~\eqref{Ca2-cone}, it is then deduced that  $u_c^*$ increases with the increases of the half cone angle $\theta$, the steady state liquid velocity before the transient process $u_0$, and the volume of the liquid column $V_l$, and the decrease of the radius of the tube bottom $r$. This conclusion is helpful in designing hydraulic components during cavitating transient processes.

\section{Large bubbles in the bulk liquid}
\label{sec:middlebubble}
As demonstrated in Section~\ref{sec:Observations}, large cavitation bubbles can appear both at the bottom of the tube and in the bulk liquid. 
Regarding to a large bubble initiated in bulk liquid with non-dimensional stand-off distance against the tube bottom $\gamma=h/R_0>0$ ($h$ is the distance between the initial bubble center and the tube bottom, and $R_0$ is the maximum equivalent radius of the bubble), since it grows more freely in both axial and radial directions, a complete liquid column separation is more difficult to occur than a bubble generated at the tube bottom, as exemplified in Fig.~\ref{sequence}(c).

\begin{figure*}[t]
\centering
\includegraphics[width=\linewidth]{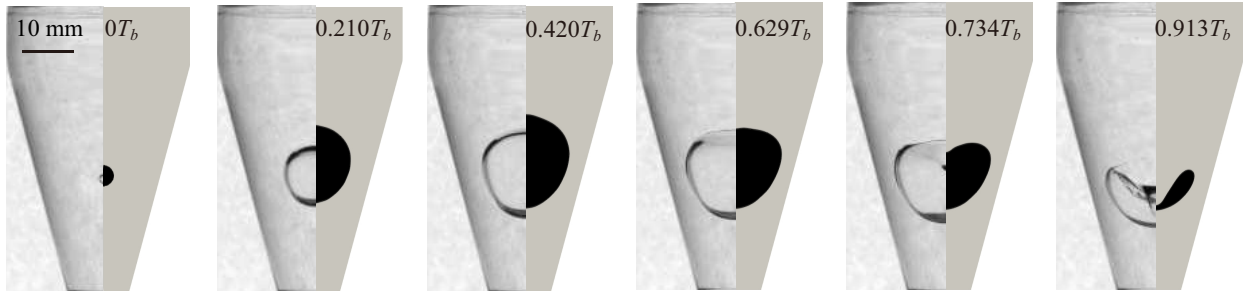}
\caption{Comparison between the simulation and the experimental case in Fig.~\ref{sequence}~(c). The left part in a single frame is the experimental result and the right part is the simulation result. The initial pressure and radius of the bubble are $p_i=1.27$~MPa, $R_i=2~\mathrm{mm}$. The height of the water column $l=550~\mathrm{mm}$. In this case, $h=22.4~\mathrm{mm}$, $R_0=9.5~\mathrm{mm}$, and $\gamma=h/R_0=2.36$, where $h$ is the standoff distance from the initial bubble center to the boundary and $R_0$ is the equivalent maximum bubble radius. $T_b$ is the duration from the bubble inception to its first collapse.}
\label{Simulation1}
\end{figure*}

Next, in order to evaluate the damaging power of large cavitation bubbles generated in the bulk liquid in the middle of the tube, we study their dynamics including oscillation periods, jet characteristics and pressure pulses upon collapses. 
We rely mostly on the numerical simulations for the discussions. 
The details of the numerical simulations are presented in Appendix \ref{Appendix}.
The numerical technique is firstly validated against the experimental case presented in Fig.~\ref{sequence}(c), as shown in Fig.~\ref{Simulation1}. 
It is seen that the main features of the bubble dynamics during the bubble growth and collapse, especially the bubble deformation and the jetting, match well with the corresponding high-speed images.
However, since the adopted numerical approach approximates the evolution of the cavitation bubble with an initially compressed air bubble, and ignores the complex mass transfer by condensation during the bubble collapse, the rebound characteristics of the simulations may deviate from the experimental results \cite{Zeng,Wu,Ohl}. 
We thus focus on the first cycle of bubble expansion and collapse for the following discussions. 

To illustrate the influence of the tubes on the dynamics of large bubbles, we first compare three cases, where the bubble locates near an infinite plate, in a straight cylindrical tube, and in a conical-frustum shaped tube section, respectively. 
The three cavitation bubbles have the same maximum equivalent radius $R_0=9.5~\mathrm{mm}$ and the same standoff distance $h=22.4~\mathrm{mm}$ from the tube bottom (or the plate), hence $\gamma=2.36$. The numerical results are analyzed below.

\begin{figure*}[t]
\centering
\includegraphics[width=\linewidth]{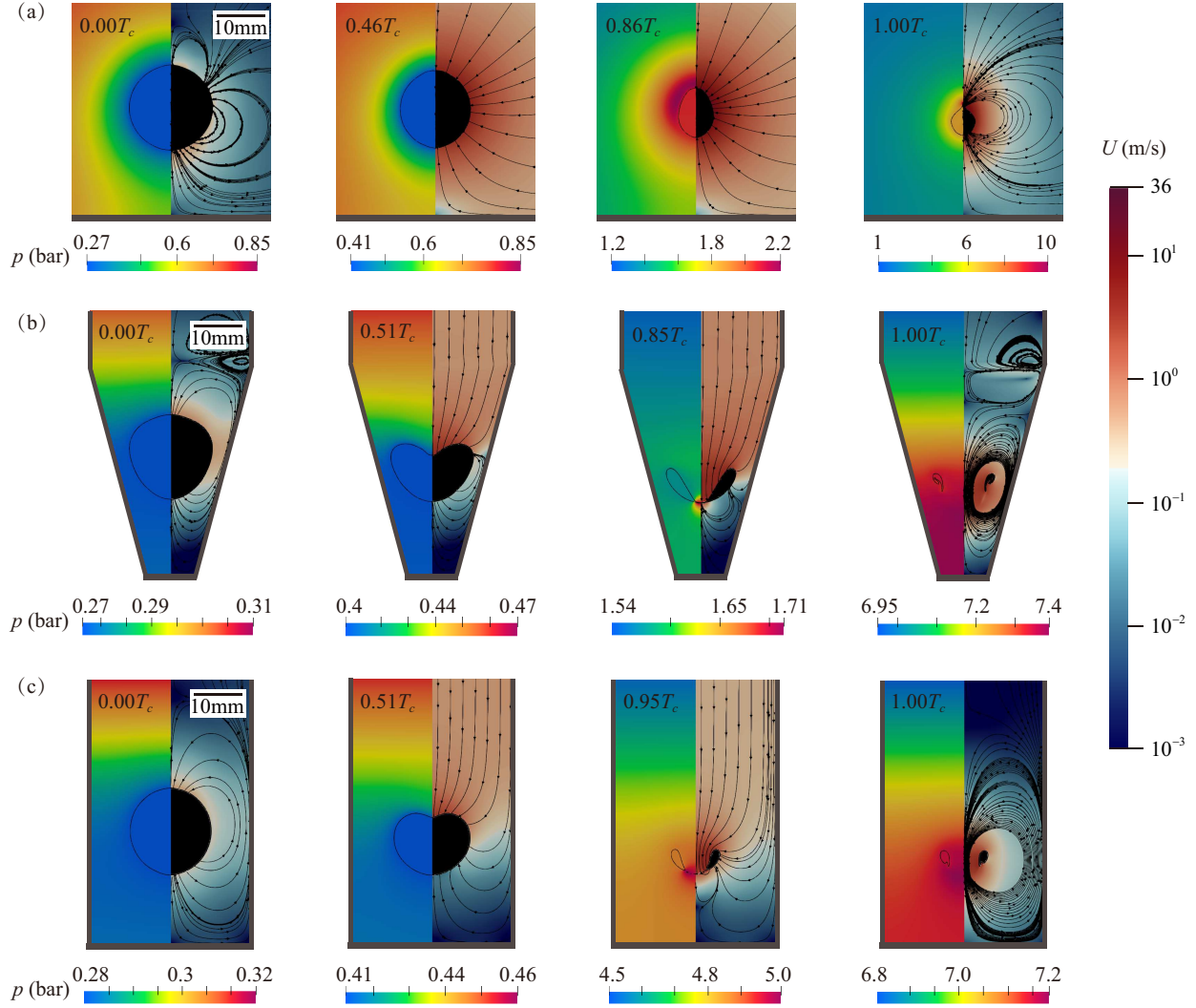}
\caption{Calculated pressure (left part in each frame) and velocity fields (right part in each frame) during collapse of a cavitation bubble (a) near an infinite rigid boundary, (b) in a tube with a conical-frustum shaped closed end, and (c) in a straight cylindrical  tube. $h=22.4~\mathrm{mm}$, $R_0=9.5~\mathrm{mm}$, i.e., $\gamma=h/R_0=2.36$. Heights of the water column $l=550~\mathrm{mm}$ for both (b) and (c).}  
\label{Simulation2}
\end{figure*}

\begin{table}[t]
\caption{\label{table2}The collapse times and jet speeds of the bubbles in three different geometries.}
\begin{ruledtabular}
\begin{tabular}{cccc}
 &Infinite plate ($\theta=90^{\circ}$)&Conical frustum ($\theta=15^{\circ}$)&Cylinder ($\theta=0^{\circ}$)\\
\hline
$T_c$ (ms)&1.26&7.93&7.93\\
$U_j$ (m/s)&60.0&3.86&3.56\\
\end{tabular}
\end{ruledtabular}
\end{table}

As summarized in Table~\ref{table2}, the characteristic collapse times $T_c$ of the bubbles in the two tubes ($7.93$~ms) are much longer than that near the infinite plate ($1.26$~ms). 
In fact, when a bubble collapses near a boundary, its collapse time can be theoretically evaluated by modifying Rayleigh equation with a function of $\gamma$ \cite{Rattray,Milton}, and approaches the Rayleigh time of a spherically collapsing bubble determined only by the maximum bubble size as $\gamma \rightarrow \infty$, given the pressure difference $\Delta p=p_\infty-p_v$ and the liquid density $\rho_l$. 
As shown in Fig.~\ref{Simulation2}(a), the flat infinite boundary has relatively limited influence on the dynamics of the bubble during most of the collapse of a bubble with $\gamma=2.36$. 
On the contrary, for large bubbles in the tubes, the tube walls have significant impact on the flow fields even when the boundary of the bubbles are not too close to the bottom of the tubes.
Liquids around the bubbles are restricted and the bubbles collapse mostly along the axial direction of the tube, as shown in Fig.~\ref{Simulation2}(b) and (c).
Recalling the quasi one-dimensional analysis in Section~\ref{sec:bottombubble}, the collapse times of these bubbles can also be approximated with Eq.~\ref{Tc-cone} and are again determined by not only the maximum bubble sizes, but also the volumes of the liquid columns. 

When the bubble in Fig.~\ref{sequence}(c) ($\gamma=2.36$) collapses, its top first flattens and forms a dimple at the center.
The dimple then develops into a jet, which eventually penetrates the bottom of the bubble. 
For such a large bubble in a long tube filled with water, the pressure field anisotropies around the bubble that break its sphericity are attributed to the combined effects of nearby boundaries \cite{Benjamin,Brennen}, gravity \cite{obreschkow2011universal}, and the asymmetry between the masses of the water column on each side of the bubble \cite{ory2000growth}. 
As a result, the jets in the conical and cylindrical tube penetrate the bottom of the bubbles at 0.85~$T_c$ and 0.95~$T_c$ before the first collapse, as shown in Fig.~\ref{Simulation2}(b) and (c). However, the bubble near the plate collapses uniformly during most of the time period of the collapse. A narrow jet is generated until at the end of collapse (as shown in Fig.~\ref{Simulation2}(a)), and grows only significantly in the rebound stage. Therefore, as classified in Ref.~\cite{Supponen} in term of the occurrence time, the jets in the tubes can be defined as ``strong jet'' and the jet near the plate is ``intermediate'' or ``weak''.
Table~\ref{table2} also summarizes the maximum jet speed $U_j$ before it penetrates the opposite bubble surface in the three cases. 
The higher $U_j$ value ($60.0$~m/s) of the bubble near the plate measures the much larger bubble collapsing velocity, comparing with the finite collapsing velocities of the bubbles in the tubes ($3.86$~m/s and $3.56$~m/s respectively). 
The velocity difference between the bubbles in the two tubes indicates that the flow focusing effects in the conical tube section slightly accelerate the downward jet. 

\begin{figure*}[t]
\centering
\includegraphics[width=\linewidth]{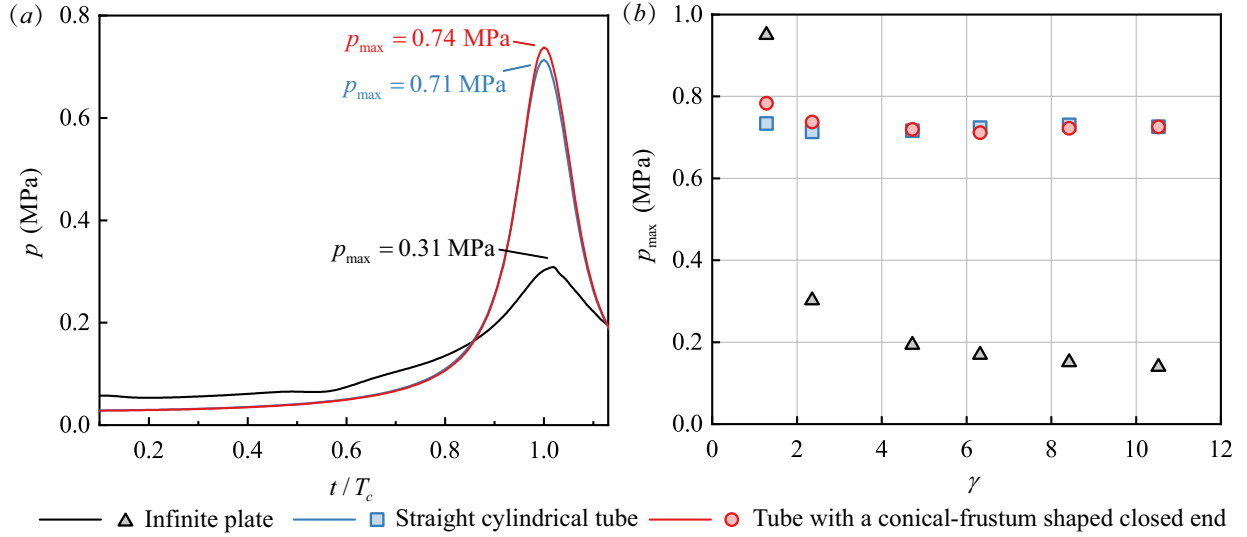}
\caption{The pressure monitored at $x=0.2$~mm on the symmetry axis in Fig.~\ref{domain} during the collapse. (a) The pressure evolution when $\gamma=2.36$, and (b) the maximum pressure $p_{\max}$ of different $\gamma$ in three geometries as in Fig.~\ref{Simulation2}.}  
\label{pressure}
\end{figure*}

The comparison of the pressure evolution at a pressure monitoring point in the flow field, $x=0.2$~mm on the symmetry axis in Fig.~\ref{domain}, is shown in Fig.~\ref{pressure}(a). 
The maximum pressure pulse during the bubble collapse $p_{\max}$ can serve as a measure of the damage power to the tube or the plate. 
It is seen that in all three cases $p_{\max}$ occurs at the moment of the collapse, and that $p_{\max}$ is much higher in tubes (0.74~MPa and 0.71~MPa respectively) than near the plate (0.31~MPa). 
A large cavitation bubble in a tube exhibits a much stronger damaging ability to the boundaries, and the conical-shaped section further increases its power.

Moreover, the influence of $\gamma$ on $p_{\max}$ for the bubbles in three different geometries are investigated. As shown in Fig.~\ref{pressure}(b), $p_{\max}$ is compared at six $\gamma$ values ($\gamma=$1.18, 2.36, 4.72, 6.32, 8.42, and 10.53). 
For the large range of $\gamma$ values covered in the tubes, $p_{\max}$ remains significant.
For a bubble very close to a plate, e.g., $\gamma=1.18$, it causes rather high pressure resulting from the high-speed jet impacting on the wall surface at collapse. However, when $\gamma$ increases, e.g., $\gamma>2.36$, the bubble collapse no longer induces substantial pressure increase near the wall.
In this sense, a large bubble in a tube is more dangerous as it consistently generates high pressure at the closed end.

\section{Conclusions}
In the current paper, we investigate the dynamics of large cavitation bubbles generated both at the tube bottom and in the bulk liquid in tubes with  conical-frustum shaped closed ends during  transient process with the tube-arrest method. 
Bubbles generated at the tube bottom are more likely to induce full liquid column separation, given the same bubble volume as the ones in the bulk liquid. 
Thus we focus on the onset criteria of large bubbles for the former cases, and the collapse time, the jet and pressure characteristics for the latter cases.  

To evaluate the conditions for the occurrence of cavitation, we modify the results of Pan \emph{et al.} \cite{Pan} in the cylindrical tube during a transient process to include the tube geometric characteristics, and derive a non-dimensional parameter $Ca_1= \frac{p_\infty}{\rho_l a V_l A^{-1}}$ referring to the quasi one-dimensional arguments. 
To evaluate the maximum size $V_{\max}$ of the bubbles generated at the tube bottom,  we derive a more universal non-dimensional parameter $Ca_2=2\tan\theta/(\sqrt[3]{\frac{1}{Ca_0}\frac{V_l}{V_0}+1}-1)$ considering energy conservation by modifying the criteria described in Ref.~\cite{Xu}. 
Thus, the onset of large conical-frustum shaped cavitation bubbles generated at tube bottom that induce liquid column separation requires both $Ca_1<1$ and $Ca_2<1$, which is validated by systematic experiments based on the tube-arrest approach.
We further propose a Rayleigh-type equation to describe the collapse of these bubbles, and reveal that the collapse time $T_c$ is related to both the maximum size of the bubble $V_{\max}$ and the volume of the liquid column $V_l$. 
Moreover, the bubble collapse speed  $u_c^*$ increases solely with $\tan \theta \cdot Ca_2^{-1}$. 
Thus, considering alleviating the damage of the cavitation during transient processes in a hydraulic system, it is favorable to have a tube with a small cone angle and a large radius of the tube closed end, a short water column, and a slow steady-state liquid flow in the tube.

For the bubbles generated in the bulk liquid in the middle of the tube, we numerically study their characteristics during collapse. 
We observe a much stronger jet of bubbles in the tubes, in terms of its time of occurrence, comparing with a bubble near an infinite plate with the same standoff distance $\gamma$ from the tube bottom or the plate. 
The maximum pressure pulse $p_{\max}$ occurs at bubble collapse and remains significant for the large range of $\gamma$ values covered in the study. 
These features indicate that large bubbles generated in the bulk liquid in the middle of the tube are much more damaging than the ones near a plate.

In summary, we conduct systematical studies on large transient cavitation bubbles in a tube, considering the effects of the tube geometry and the bubble location. 
Our research can provide guidelines in the design and safe operation of hydraulic machinery with complex geometries during the transient processes, for example, the rapid closing of valves or guide vanes in the upstream of the draft tube of a hydraulic turbine.

\section*{Supplementary material}
See supplementary material for the movie of an example of the cavitation bubble damaging the tube bottom in our experiments.

\begin{acknowledgments}

The  work  was supported  by  National  Natural  Science  Foundation  of  China  (No. 52076120 and 52079066), the Academic Research Projects of Beijing Union University (No. ZK90202108), the State Key Laboratory of Hydroscience and Engineering (2019-KY-04, sklhse-2019-E-02, sklhse-2020-E-03 and 2020-KY-01), the Creative Seed Fund of Shanxi Research Institute for Clean Energy, Tsinghua University, and the National Experimental Teaching Demonstration Center for Power Engineering and Engineering Thermophysics, Tsinghua University.

\end{acknowledgments}

\appendix
\section{Details of numerical simulation}
\label{Appendix}

The numerical method is adapted from Zeng \emph{et al.} \cite{Zeng}, and is briefly introduced below. Simulations are conducted using OpenFOAM. A cavitation bubble is modeled as a non-condensable air bubble starting from a small radius $R_i=2$~mm with a high internal pressure $p_i=1.27$~MPa in the beginning for all our cases.

Ignoring the mass transfer and thermal effects, the flow field is resolved by solving compressible Navier-Stokes equations 
\begin{align*}
  \frac{\partial \rho}{\partial t}+\nabla \cdot(\rho \bm{u})=0,
\end{align*}
\begin{align*}
  \frac{\partial \rho \bm{u}}{\partial t}+\nabla \cdot(\rho \bm{u}\bm{u})=-\nabla p+\nabla \cdot \bm{s}+\rho \bm{g}+\bm{f},
\end{align*}
where $\rho=\alpha \rho_l+(1-\alpha)\rho_g$ is the fluid density ($0\le\alpha\le1$ is the volume fraction of the liquid phase, and subscripts `$l$' and `$g$' represent respectively liquid and gas phases hereinafter. $\bm{g}$ is the gravitational acceleration. $\bm{f}$ is the source term due to surface tension, which is modelled with the continuous-surface-force (CSF) method. $\bm{s}=\mu(\nabla \bm{u}+\nabla \bm{u}^{\rm{T}}-2/3(\nabla \cdot \bm{u})\bm{I}$ is the viscous stress tensor, in which $\bm{I}$ is the identity tensor and  $\mu=\alpha \mu_l+(1-\alpha)\mu_g$ is the dynamic viscosity.

The gas-water interface is captured with the volume-of-fluid (VOF) method \cite{Zeng}. The transport equation of the volume fraction of the liquid phase $\alpha$ reads
\begin{align*}
  \frac{\partial \alpha}{\partial t}+ \nabla \cdot (\alpha \bm{u})+  \nabla \cdot (\alpha(1-\alpha)\bm{U_r})=\alpha(1-\alpha)(\frac{\psi_g}{\rho_g}-\frac{\psi_l}{\rho_l})\frac{\rm{D}p}{\rm{D}t}  
 + \alpha \nabla \cdot \bm{u},
\end{align*} 
where  $u$ is the velocity field,  $\bm{U_r}$ is the relative velocity between the gas and liquid phases, and $\psi=\rm{D}\rho/\rm{D}p$ for both gas and liquid phases.

\begin{figure}[t]
\centering
\includegraphics[width=3in]{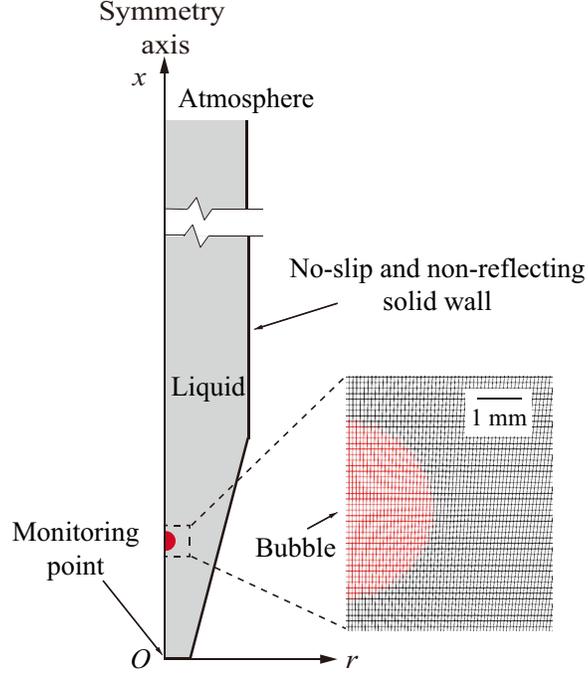}
\caption{Computation domain and the mesh.}
\label{domain}
\end{figure}

The equation of state (EoS) for the liquid is
\begin{align*}
  \rho_l=\rho_{l0}\left(\frac{p+B}{p_{l0}+B}\right)^{\frac{1}{\gamma_l}},
\end{align*}
where $\rho_{l0}=998.2~\mathrm{kg/m^3}$ and $p_{l0}=101,325~\mathrm{Pa}$ are the reference density and the reference pressure respectively. $\gamma_l$ and $B$ are set to be 7.15 and $3.04\times10^8~\mathrm{Pa}$ respectively.

For the gas, the adiabatic equation of state is adopted
\begin{align*}
  \rho_g=\rho_{g0}\left(\frac{p}{p_{g0}}\right)^{\frac{1}{\gamma_g}},
\end{align*}
where $\gamma_g=1.0$ is the specific heat ratio and $\rho_{g0}=0.12~\mathrm{kg/m^3}$ is the reference gas density under reference pressure $p_{g0}=101,325~\mathrm{Pa}$.

\begin{figure}[t]
\centering
\includegraphics[width=3.5in]{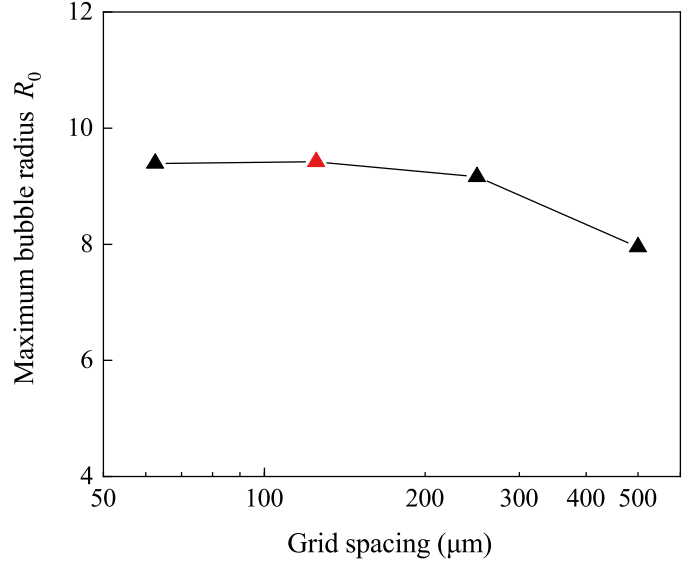}
\caption{The grid independence test using the equivalent maximum bubble radius $R_0$. The red symbol represents the grid spacing we adopt in the simulation.}
\label{Gridindependence}
\end{figure}

Taking advantage of that the bubble grows and collapses in an axisymmetric fashion as shown in Fig.~\ref{sequence}(a)-(c), we define a wedge-shaped ($1 ^\circ$ in circumferential direction) computational domain to reduce the calculation load. 
The sectional view of the computational domain and the initial mesh system are shown in Fig.~\ref{domain}. 
For the validation case, the domain is 5~--~17~mm in radial (17 grid points) and 550~mm (550 grid points) in axial directions referring to the experiment. In the area of interest, i.e., where the bubble is located, we specially refine the mesh and perform the grid independence test of the corresponding case in Fig.~\ref{sequence}(c). As shown in Fig.~\ref{Gridindependence}, the equivalent maximum bubble radius $R_0$ shows good convergence when the grid spacing $\Delta x \le$250~~$\mathrm{\mu m}$, and the grid spacing 125~~$\mathrm{\mu m}$ is adopted in our simulation.

A uniform constant atmospheric pressure (101,325~Pa) is imposed on the top boundary (free surface), and  no-slip and non-reflecting boundary conditions are used at the rigid boundaries \cite{Toro,Trummler}.  The gravitational acceleration and the surface tension coefficient of the water-air interface are 9.8~$\mathrm{m/s^2}$  and 0.07~N/m respectively. The adjustable time step  $\Delta t$ is chosen to satisfy that the Courant number $u\Delta t/\Delta x\le0.1$, where $\Delta x$ is the size of the mesh.

\providecommand{\noopsort}[1]{}\providecommand{\singleletter}[1]{#1}%


\begin{thebibliography}{36}%
\makeatletter
\providecommand \@ifxundefined [1]{%
 \@ifx{#1\undefined}
}%
\providecommand \@ifnum [1]{%
 \ifnum #1\expandafter \@firstoftwo
 \else \expandafter \@secondoftwo
 \fi
}%
\providecommand \@ifx [1]{%
 \ifx #1\expandafter \@firstoftwo
 \else \expandafter \@secondoftwo
 \fi
}%
\providecommand \natexlab [1]{#1}%
\providecommand \enquote  [1]{``#1''}%
\providecommand \bibnamefont  [1]{#1}%
\providecommand \bibfnamefont [1]{#1}%
\providecommand \citenamefont [1]{#1}%
\providecommand \href@noop [0]{\@secondoftwo}%
\providecommand \href [0]{\begingroup \@sanitize@url \@href}%
\providecommand \@href[1]{\@@startlink{#1}\@@href}%
\providecommand \@@href[1]{\endgroup#1\@@endlink}%
\providecommand \@sanitize@url [0]{\catcode `\\12\catcode `\$12\catcode
  `\&12\catcode `\#12\catcode `\^12\catcode `\_12\catcode `\%12\relax}%
\providecommand \@@startlink[1]{}%
\providecommand \@@endlink[0]{}%
\providecommand \url  [0]{\begingroup\@sanitize@url \@url }%
\providecommand \@url [1]{\endgroup\@href {#1}{\urlprefix }}%
\providecommand \urlprefix  [0]{URL }%
\providecommand \Eprint [0]{\href }%
\providecommand \doibase [0]{http://dx.doi.org/}%
\providecommand \selectlanguage [0]{\@gobble}%
\providecommand \bibinfo  [0]{\@secondoftwo}%
\providecommand \bibfield  [0]{\@secondoftwo}%
\providecommand \translation [1]{[#1]}%
\providecommand \BibitemOpen [0]{}%
\providecommand \bibitemStop [0]{}%
\providecommand \bibitemNoStop [0]{.\EOS\space}%
\providecommand \EOS [0]{\spacefactor3000\relax}%
\providecommand \BibitemShut  [1]{\csname bibitem#1\endcsname}%
\let\auto@bib@innerbib\@empty
%</preamble>
\bibitem [{\citenamefont {Nishi}\ and\ \citenamefont {Liu}(2013)}]{Nishi}%
  \BibitemOpen
  \bibfield  {author} {\bibinfo {author} {\bibfnamefont {M.}~\bibnamefont
  {Nishi}}\ and\ \bibinfo {author} {\bibfnamefont {S.}~\bibnamefont {Liu}},\
  }\bibfield  {title} {\enquote {\bibinfo {title} {An outlook on the
  draft-tube-surge study},}\ }\href {\doibase 10.5293/ijfms.2013.6.1.033}
  {\bibfield  {journal} {\bibinfo  {journal} {Int. J. Fluid Mach. Syst.}\
  }\textbf {\bibinfo {volume} {6}},\ \bibinfo {pages} {33--48} (\bibinfo {year}
  {2013})}\BibitemShut {NoStop}%
\bibitem [{\citenamefont {Mishra}\ and\ \citenamefont {Peles}(2006)}]{Mishra}%
  \BibitemOpen
  \bibfield  {author} {\bibinfo {author} {\bibfnamefont {C.}~\bibnamefont
  {Mishra}}\ and\ \bibinfo {author} {\bibfnamefont {Y.}~\bibnamefont {Peles}},\
  }\bibfield  {title} {\enquote {\bibinfo {title} {An experimental
  investigation of hydrodynamic cavitation in micro-venturis},}\ }\href
  {\doibase 10.1063/1.2360996} {\bibfield  {journal} {\bibinfo  {journal}
  {Phys. Fluids}\ }\textbf {\bibinfo {volume} {18}},\ \bibinfo {pages} {103603}
  (\bibinfo {year} {2006})}\BibitemShut {NoStop}%
\bibitem [{\citenamefont {Barre}\ \emph {et~al.}(2009)\citenamefont {Barre},
  \citenamefont {Rolland}, \citenamefont {Boitel}, \citenamefont {Goncalves},\
  and\ \citenamefont {Fortes~Patella}}]{Barre}%
  \BibitemOpen
  \bibfield  {author} {\bibinfo {author} {\bibfnamefont {S.}~\bibnamefont
  {Barre}}, \bibinfo {author} {\bibfnamefont {J.}~\bibnamefont {Rolland}},
  \bibinfo {author} {\bibfnamefont {G.}~\bibnamefont {Boitel}}, \bibinfo
  {author} {\bibfnamefont {E.}~\bibnamefont {Goncalves}}, \ and\ \bibinfo
  {author} {\bibfnamefont {R.}~\bibnamefont {Fortes~Patella}},\ }\bibfield
  {title} {\enquote {\bibinfo {title} {Experiments and modeling of cavitating
  flows in venturi: attached sheet cavitation},}\ }\href {\doibase
  10.1016/j.euromechflu.2008.09.001} {\bibfield  {journal} {\bibinfo  {journal}
  {Eur. J. Mech. B/Fluids}\ }\textbf {\bibinfo {volume} {28}},\ \bibinfo
  {pages} {444--464} (\bibinfo {year} {2009})}\BibitemShut {NoStop}%
\bibitem [{\citenamefont {Zhang}\ \emph {et~al.}(2019)\citenamefont {Zhang},
  \citenamefont {Zuo}, \citenamefont {M{\o}rch},\ and\ \citenamefont
  {Liu}}]{Haochen}%
  \BibitemOpen
  \bibfield  {author} {\bibinfo {author} {\bibfnamefont {H.}~\bibnamefont
  {Zhang}}, \bibinfo {author} {\bibfnamefont {Z.}~\bibnamefont {Zuo}}, \bibinfo
  {author} {\bibfnamefont {K.~A.}\ \bibnamefont {M{\o}rch}}, \ and\ \bibinfo
  {author} {\bibfnamefont {S.}~\bibnamefont {Liu}},\ }\bibfield  {title}
  {\enquote {\bibinfo {title} {Thermodynamic effects on venturi cavitation
  characteristics},}\ }\href {\doibase 10.1063/1.5116156} {\bibfield  {journal}
  {\bibinfo  {journal} {Phys. Fluids}\ }\textbf {\bibinfo {volume} {31}},\
  \bibinfo {pages} {097107} (\bibinfo {year} {2019})}\BibitemShut {NoStop}%
\bibitem [{\citenamefont {Bonin}(1960)}]{Bonin}%
  \BibitemOpen
  \bibfield  {author} {\bibinfo {author} {\bibfnamefont {C.}~\bibnamefont
  {Bonin}},\ }\bibfield  {title} {\enquote {\bibinfo {title} {Water-hammer
  damage to oigawa power station},}\ }\href {\doibase 10.1115/1.3672721}
  {\bibfield  {journal} {\bibinfo  {journal} {ASME J. Eng. Power}\ }\textbf
  {\bibinfo {volume} {82}},\ \bibinfo {pages} {111} (\bibinfo {year}
  {1960})}\BibitemShut {NoStop}%
\bibitem [{\citenamefont {Bergant}, \citenamefont {Simpson},\ and\
  \citenamefont {Tijsseling}(2006)}]{Bergant2006}%
  \BibitemOpen
  \bibfield  {author} {\bibinfo {author} {\bibfnamefont {A.}~\bibnamefont
  {Bergant}}, \bibinfo {author} {\bibfnamefont {A.~R.}\ \bibnamefont
  {Simpson}}, \ and\ \bibinfo {author} {\bibfnamefont {A.~S.}\ \bibnamefont
  {Tijsseling}},\ }\bibfield  {title} {\enquote {\bibinfo {title} {Water hammer
  with column separation: A historical review},}\ }\href {\doibase
  10.1016/j.jfluidstructs.2005.08.008} {\bibfield  {journal} {\bibinfo
  {journal} {J. Fluids Struct.}\ }\textbf {\bibinfo {volume} {22}},\ \bibinfo
  {pages} {135--171} (\bibinfo {year} {2006})}\BibitemShut {NoStop}%
\bibitem [{\citenamefont {Nonoshita}, \citenamefont {Matsumoto},\ and\
  \citenamefont {Kubota}(1999)}]{Nonoshita}%
  \BibitemOpen
  \bibfield  {author} {\bibinfo {author} {\bibfnamefont {T.}~\bibnamefont
  {Nonoshita}}, \bibinfo {author} {\bibfnamefont {H.}~\bibnamefont {Matsumoto},
  \bibfnamefont {Y.and~Ohashi}}, \ and\ \bibinfo {author} {\bibfnamefont
  {T.}~\bibnamefont {Kubota}},\ }\bibfield  {title} {\enquote {\bibinfo {title}
  {Water column separation in a straight draft tube},}\ }in\ \href@noop {}
  {\emph {\bibinfo {booktitle} {Proceedings of the Third ASME-JSME Joint Fluids
  Engineering Conference}}}\ (\bibinfo {address} {San Francisco, USA},\
  \bibinfo {year} {1999})\BibitemShut {NoStop}%
\bibitem [{\citenamefont {Pejovic}, \citenamefont {Karney},\ and\ \citenamefont
  {Zhang}(2004)}]{Pejovic2004}%
  \BibitemOpen
  \bibfield  {author} {\bibinfo {author} {\bibfnamefont {S.}~\bibnamefont
  {Pejovic}}, \bibinfo {author} {\bibfnamefont {B.}~\bibnamefont {Karney}}, \
  and\ \bibinfo {author} {\bibfnamefont {Q.}~\bibnamefont {Zhang}},\ }\bibfield
   {title} {\enquote {\bibinfo {title} {Water column separation in long
  tailrace tunnel},}\ }in\ \href@noop {} {\emph {\bibinfo {booktitle}
  {Hydroturbo 2004, International Conference on Hydro-Power Engineering}}}\
  (\bibinfo {address} {Brno, Czech Republic},\ \bibinfo {year} {2004})\ pp.\
  \bibinfo {pages} {18--22}\BibitemShut {NoStop}%
\bibitem [{\citenamefont {Pejovic}, \citenamefont {Karney},\ and\ \citenamefont
  {Gajic}(2011)}]{Pejovic2011}%
  \BibitemOpen
  \bibfield  {author} {\bibinfo {author} {\bibfnamefont {S.}~\bibnamefont
  {Pejovic}}, \bibinfo {author} {\bibfnamefont {B.}~\bibnamefont {Karney}}, \
  and\ \bibinfo {author} {\bibfnamefont {A.}~\bibnamefont {Gajic}},\ }\bibfield
   {title} {\enquote {\bibinfo {title} {Analysis of pump-turbine s instability
  and reverse waterhammer incidents in hydropower systems},}\ }in\ \href@noop
  {} {\emph {\bibinfo {booktitle} {4th International Meeting on Cavitation and
  Dynamic Problems in Hydraulic Machinery Systems}}}\ (\bibinfo {address}
  {Belgrade, Serbia},\ \bibinfo {year} {2011})\BibitemShut {NoStop}%
\bibitem [{\citenamefont {Zhang}\ \emph {et~al.}(2016)\citenamefont {Zhang},
  \citenamefont {Cheng}, \citenamefont {Xia},\ and\ \citenamefont
  {Yang}}]{Zhang}%
  \BibitemOpen
  \bibfield  {author} {\bibinfo {author} {\bibfnamefont {X.}~\bibnamefont
  {Zhang}}, \bibinfo {author} {\bibfnamefont {Y.}~\bibnamefont {Cheng}},
  \bibinfo {author} {\bibfnamefont {L.}~\bibnamefont {Xia}}, \ and\ \bibinfo
  {author} {\bibfnamefont {J.}~\bibnamefont {Yang}},\ }\bibfield  {title}
  {\enquote {\bibinfo {title} {{CFD} simulation of reverse water-hammer induced
  by collapse of draft-tube cavity in a model pump-turbine during runaway
  process},}\ }\href {\doibase 10.1088/1755-1315/49/5/052017} {\bibfield
  {journal} {\bibinfo  {journal} {IOP Conf.}\ }\textbf {\bibinfo {volume}
  {49}},\ \bibinfo {pages} {052017} (\bibinfo {year} {2016})}\BibitemShut
  {NoStop}%
\bibitem [{\citenamefont {He}\ \emph {et~al.}(2022)\citenamefont {He},
  \citenamefont {Yang}, \citenamefont {Yang}, \citenamefont {Hu},\ and\
  \citenamefont {Peng}}]{He}%
  \BibitemOpen
  \bibfield  {author} {\bibinfo {author} {\bibfnamefont {X.}~\bibnamefont
  {He}}, \bibinfo {author} {\bibfnamefont {J.}~\bibnamefont {Yang}}, \bibinfo
  {author} {\bibfnamefont {J.}~\bibnamefont {Yang}}, \bibinfo {author}
  {\bibfnamefont {J.}~\bibnamefont {Hu}}, \ and\ \bibinfo {author}
  {\bibfnamefont {T.}~\bibnamefont {Peng}},\ }\bibfield  {title} {\enquote
  {\bibinfo {title} {Experimental study of cavitating vortex rope and water
  column separation in a pump turbine},}\ }\href {\doibase 10.1063/5.0086509}
  {\bibfield  {journal} {\bibinfo  {journal} {Phys. Fluids}\ }\textbf {\bibinfo
  {volume} {34}},\ \bibinfo {pages} {044101} (\bibinfo {year}
  {2022})}\BibitemShut {NoStop}%
\bibitem [{\citenamefont {Sharp}(1960)}]{Sharp}%
  \BibitemOpen
  \bibfield  {author} {\bibinfo {author} {\bibfnamefont {B.~B.}\ \bibnamefont
  {Sharp}},\ }\bibfield  {title} {\enquote {\bibinfo {title} {Cavity formation
  in simple pipes due to rupture of the water column},}\ }\href {\doibase
  10.1038/185302b0} {\bibfield  {journal} {\bibinfo  {journal} {Nature}\
  }\textbf {\bibinfo {volume} {185}},\ \bibinfo {pages} {302--303} (\bibinfo
  {year} {1960})}\BibitemShut {NoStop}%
\bibitem [{\citenamefont {Simpson}\ and\ \citenamefont
  {Wylie}(1991)}]{Simpson}%
  \BibitemOpen
  \bibfield  {author} {\bibinfo {author} {\bibfnamefont {A.~R.}\ \bibnamefont
  {Simpson}}\ and\ \bibinfo {author} {\bibfnamefont {E.~B.}\ \bibnamefont
  {Wylie}},\ }\bibfield  {title} {\enquote {\bibinfo {title} {Large
  water-hammer pressures for column separation in pipelines},}\ }\href
  {\doibase 10.1061/(ASCE)0733-9429(1991)117:10(1310)} {\bibfield  {journal}
  {\bibinfo  {journal} {ASCE J. Hydraul. Eng.}\ }\textbf {\bibinfo {volume}
  {117}},\ \bibinfo {pages} {1310--1316} (\bibinfo {year} {1991})}\BibitemShut
  {NoStop}%
\bibitem [{\citenamefont {Brunone}, \citenamefont {Golia},\ and\ \citenamefont
  {Greco}(1991)}]{Brunone}%
  \BibitemOpen
  \bibfield  {author} {\bibinfo {author} {\bibfnamefont {B.}~\bibnamefont
  {Brunone}}, \bibinfo {author} {\bibfnamefont {U.}~\bibnamefont {Golia}}, \
  and\ \bibinfo {author} {\bibfnamefont {M.}~\bibnamefont {Greco}},\ }\bibfield
   {title} {\enquote {\bibinfo {title} {Modelling of fast transients by
  numerical methods},}\ }in\ \href@noop {} {\emph {\bibinfo {booktitle}
  {Proceedings of the International Meeting on Hydraulic Transients with Column
  Separation}}}\ (\bibinfo {address} {Valencia, Spain},\ \bibinfo {year}
  {1991})\ pp.\ \bibinfo {pages} {215--222}\BibitemShut {NoStop}%
\bibitem [{\citenamefont {Bergant}\ and\ \citenamefont
  {Simpson}(1999)}]{Bergant1999}%
  \BibitemOpen
  \bibfield  {author} {\bibinfo {author} {\bibfnamefont {A.}~\bibnamefont
  {Bergant}}\ and\ \bibinfo {author} {\bibfnamefont {A.~R.}\ \bibnamefont
  {Simpson}},\ }\bibfield  {title} {\enquote {\bibinfo {title} {Pipeline column
  separation flow regimes},}\ }\href {\doibase
  10.1061/(ASCE)0733-9429(1999)125:8(835)} {\bibfield  {journal} {\bibinfo
  {journal} {ASCE J. Hydraul. Eng.}\ }\textbf {\bibinfo {volume} {125}},\
  \bibinfo {pages} {835--848} (\bibinfo {year} {1999})}\BibitemShut {NoStop}%
\bibitem [{\citenamefont {Adamkowski}\ and\ \citenamefont
  {Lewandowski}(2012)}]{Adamkowski2012}%
  \BibitemOpen
  \bibfield  {author} {\bibinfo {author} {\bibfnamefont {A.}~\bibnamefont
  {Adamkowski}}\ and\ \bibinfo {author} {\bibfnamefont {M.}~\bibnamefont
  {Lewandowski}},\ }\bibfield  {title} {\enquote {\bibinfo {title}
  {Investigation of hydraulic transients in a pipeline with column
  separation},}\ }\href {\doibase 10.1061/(ASCE)HY.1943-7900.0000596}
  {\bibfield  {journal} {\bibinfo  {journal} {ASCE J. Hydrual. ENG.}\ }\textbf
  {\bibinfo {volume} {138}},\ \bibinfo {pages} {935--944} (\bibinfo {year}
  {2012})}\BibitemShut {NoStop}%
\bibitem [{\citenamefont {Adamkowski}\ and\ \citenamefont
  {Lewandowski}(2015)}]{Adamkowski2015}%
  \BibitemOpen
  \bibfield  {author} {\bibinfo {author} {\bibfnamefont {A.}~\bibnamefont
  {Adamkowski}}\ and\ \bibinfo {author} {\bibfnamefont {M.}~\bibnamefont
  {Lewandowski}},\ }\bibfield  {title} {\enquote {\bibinfo {title} {Cavitation
  characteristics of shutoff valves in numerical modeling of transients in
  pipelines with column separation},}\ }\href {\doibase
  10.1061/(ASCE)HY.1943-7900.0000971} {\bibfield  {journal} {\bibinfo
  {journal} {ASCE J. Hydraul. Eng.}\ }\textbf {\bibinfo {volume} {141}},\
  \bibinfo {pages} {04014077} (\bibinfo {year} {2015})}\BibitemShut {NoStop}%
\bibitem [{\citenamefont {Daily}\ \emph {et~al.}(2014)\citenamefont {Daily},
  \citenamefont {Pendlebury}, \citenamefont {Langley}, \citenamefont {Hurd},
  \citenamefont {Thomson},\ and\ \citenamefont {Truscott}}]{Jesse}%
  \BibitemOpen
  \bibfield  {author} {\bibinfo {author} {\bibfnamefont {J.}~\bibnamefont
  {Daily}}, \bibinfo {author} {\bibfnamefont {J.}~\bibnamefont {Pendlebury}},
  \bibinfo {author} {\bibfnamefont {K.}~\bibnamefont {Langley}}, \bibinfo
  {author} {\bibfnamefont {R.}~\bibnamefont {Hurd}}, \bibinfo {author}
  {\bibfnamefont {S.}~\bibnamefont {Thomson}}, \ and\ \bibinfo {author}
  {\bibfnamefont {T.}~\bibnamefont {Truscott}},\ }\bibfield  {title} {\enquote
  {\bibinfo {title} {Catastrophic cracking courtesy of quiescent cavitation},}\
  }\href {\doibase 10.1063/1.4894073} {\bibfield  {journal} {\bibinfo
  {journal} {Phys. Fluids}\ }\textbf {\bibinfo {volume} {26}},\ \bibinfo
  {pages} {091107} (\bibinfo {year} {2014})}\BibitemShut {NoStop}%
\bibitem [{\citenamefont {Chesterman}(1952)}]{Chesterman}%
  \BibitemOpen
  \bibfield  {author} {\bibinfo {author} {\bibfnamefont {W.~D.}\ \bibnamefont
  {Chesterman}},\ }\bibfield  {title} {\enquote {\bibinfo {title} {The dynamics
  of small transient cavities},}\ }\href {\doibase 10.1088/0370-1301/65/11/302}
  {\bibfield  {journal} {\bibinfo  {journal} {Proc. Phys. Soc. B}\ }\textbf
  {\bibinfo {volume} {65}},\ \bibinfo {pages} {846--858} (\bibinfo {year}
  {1952})}\BibitemShut {NoStop}%
\bibitem [{\citenamefont {Chen}\ and\ \citenamefont {Wang}(2004)}]{Chen}%
  \BibitemOpen
  \bibfield  {author} {\bibinfo {author} {\bibfnamefont {Q.}~\bibnamefont
  {Chen}}\ and\ \bibinfo {author} {\bibfnamefont {L.}~\bibnamefont {Wang}},\
  }\bibfield  {title} {\enquote {\bibinfo {title} {Production of large size
  single transient cavitation bubbles with tube arrest method},}\ }\href
  {\doibase 10.1088/1009-1963/13/4/028} {\bibfield  {journal} {\bibinfo
  {journal} {Chin. Phys.}\ }\textbf {\bibinfo {volume} {13}},\ \bibinfo {pages}
  {564} (\bibinfo {year} {2004})}\BibitemShut {NoStop}%
\bibitem [{\citenamefont {Xu}\ \emph {et~al.}(2021)\citenamefont {Xu},
  \citenamefont {Liu}, \citenamefont {Zuo},\ and\ \citenamefont {Pan}}]{Xu}%
  \BibitemOpen
  \bibfield  {author} {\bibinfo {author} {\bibfnamefont {P.}~\bibnamefont
  {Xu}}, \bibinfo {author} {\bibfnamefont {S.}~\bibnamefont {Liu}}, \bibinfo
  {author} {\bibfnamefont {Z.}~\bibnamefont {Zuo}}, \ and\ \bibinfo {author}
  {\bibfnamefont {Z.}~\bibnamefont {Pan}},\ }\bibfield  {title} {\enquote
  {\bibinfo {title} {On the criteria of large cavitation bubbles in a tube
  during a transient process},}\ }\href {\doibase 10.1017/jfm.2021.114}
  {\bibfield  {journal} {\bibinfo  {journal} {J. Fluid Mech.}\ }\textbf
  {\bibinfo {volume} {913}},\ \bibinfo {pages} {R6} (\bibinfo {year}
  {2021})}\BibitemShut {NoStop}%
\bibitem [{\citenamefont {Benjamin}\ and\ \citenamefont
  {Ellis}(1966)}]{Benjamin}%
  \BibitemOpen
  \bibfield  {author} {\bibinfo {author} {\bibfnamefont {T.~B.}\ \bibnamefont
  {Benjamin}}\ and\ \bibinfo {author} {\bibfnamefont {A.~T.}\ \bibnamefont
  {Ellis}},\ }\bibfield  {title} {\enquote {\bibinfo {title} {The collapse of
  cavitation bubbles and the pressures thereby produced against solid
  boundaries},}\ }\href {\doibase 10.1098/rsta.1966.0046} {\bibfield  {journal}
  {\bibinfo  {journal} {Phil. Trans. R. Soc. Lond. A}\ }\textbf {\bibinfo
  {volume} {260}},\ \bibinfo {pages} {221--240} (\bibinfo {year}
  {1966})}\BibitemShut {NoStop}%
\bibitem [{\citenamefont {Philipp}\ and\ \citenamefont
  {Lauterborn}(1998)}]{Philipp}%
  \BibitemOpen
  \bibfield  {author} {\bibinfo {author} {\bibfnamefont {A.}~\bibnamefont
  {Philipp}}\ and\ \bibinfo {author} {\bibfnamefont {W.}~\bibnamefont
  {Lauterborn}},\ }\bibfield  {title} {\enquote {\bibinfo {title} {Cavitation
  erosion by single laser-produced bubbles},}\ }\href {\doibase
  10.1017/S0022112098008738} {\bibfield  {journal} {\bibinfo  {journal} {J.
  Fluid Mech.}\ }\textbf {\bibinfo {volume} {361}},\ \bibinfo {pages} {75--116}
  (\bibinfo {year} {1998})}\BibitemShut {NoStop}%
\bibitem [{\citenamefont {Pan}\ \emph {et~al.}(2017)\citenamefont {Pan},
  \citenamefont {Kiyama}, \citenamefont {Tagawa}, \citenamefont {Daily},\ and\
  \citenamefont {Truscott}}]{Pan}%
  \BibitemOpen
  \bibfield  {author} {\bibinfo {author} {\bibfnamefont {Z.}~\bibnamefont
  {Pan}}, \bibinfo {author} {\bibfnamefont {A.}~\bibnamefont {Kiyama}},
  \bibinfo {author} {\bibfnamefont {Y.}~\bibnamefont {Tagawa}}, \bibinfo
  {author} {\bibfnamefont {D.~J.}\ \bibnamefont {Daily}}, \ and\ \bibinfo
  {author} {\bibfnamefont {T.~T.}\ \bibnamefont {Truscott}},\ }\bibfield
  {title} {\enquote {\bibinfo {title} {Cavitation onset caused by
  acceleration},}\ }\href {\doibase 10.1073/pnas.1702502114} {\bibfield
  {journal} {\bibinfo  {journal} {Proc. Natl Acad. Sci. USA}\ }\textbf
  {\bibinfo {volume} {114}},\ \bibinfo {pages} {8470--8474} (\bibinfo {year}
  {2017})}\BibitemShut {NoStop}%
\bibitem [{\citenamefont {Onuki}, \citenamefont {Oi},\ and\ \citenamefont
  {Tagawa}(2018)}]{Onuki}%
  \BibitemOpen
  \bibfield  {author} {\bibinfo {author} {\bibfnamefont {H.}~\bibnamefont
  {Onuki}}, \bibinfo {author} {\bibfnamefont {Y.}~\bibnamefont {Oi}}, \ and\
  \bibinfo {author} {\bibfnamefont {Y.}~\bibnamefont {Tagawa}},\ }\bibfield
  {title} {\enquote {\bibinfo {title} {Microjet generator for highly viscous
  fluids},}\ }\href {\doibase 10.1103/PhysRevApplied.9.014035} {\bibfield
  {journal} {\bibinfo  {journal} {Phys. Rev. Applied}\ }\textbf {\bibinfo
  {volume} {9}},\ \bibinfo {pages} {014035} (\bibinfo {year}
  {2018})}\BibitemShut {NoStop}%
\bibitem [{\citenamefont {Zeng}\ \emph {et~al.}(2018)\citenamefont {Zeng},
  \citenamefont {Gonzalez-Avila}, \citenamefont {Dijkink}, \citenamefont
  {Koukouvinis}, \citenamefont {Gavaises},\ and\ \citenamefont {Ohl}}]{Zeng}%
  \BibitemOpen
  \bibfield  {author} {\bibinfo {author} {\bibfnamefont {Q.}~\bibnamefont
  {Zeng}}, \bibinfo {author} {\bibfnamefont {S.~R.}\ \bibnamefont
  {Gonzalez-Avila}}, \bibinfo {author} {\bibfnamefont {R.}~\bibnamefont
  {Dijkink}}, \bibinfo {author} {\bibfnamefont {P.}~\bibnamefont
  {Koukouvinis}}, \bibinfo {author} {\bibfnamefont {M.}~\bibnamefont
  {Gavaises}}, \ and\ \bibinfo {author} {\bibfnamefont {C.-D.}\ \bibnamefont
  {Ohl}},\ }\bibfield  {title} {\enquote {\bibinfo {title} {Wall shear stress
  from jetting caivtation bubbles},}\ }\href {\doibase 10.1017/jfm.2018.286}
  {\bibfield  {journal} {\bibinfo  {journal} {J. Fluid Mech.}\ }\textbf
  {\bibinfo {volume} {846}},\ \bibinfo {pages} {341--355} (\bibinfo {year}
  {2018})}\BibitemShut {NoStop}%
\bibitem [{\citenamefont {Wu}\ \emph {et~al.}(2021)\citenamefont {Wu},
  \citenamefont {Li}, \citenamefont {Zuo},\ and\ \citenamefont {Liu}}]{Wu}%
  \BibitemOpen
  \bibfield  {author} {\bibinfo {author} {\bibfnamefont {S.}~\bibnamefont
  {Wu}}, \bibinfo {author} {\bibfnamefont {B.}~\bibnamefont {Li}}, \bibinfo
  {author} {\bibfnamefont {Z.}~\bibnamefont {Zuo}}, \ and\ \bibinfo {author}
  {\bibfnamefont {S.}~\bibnamefont {Liu}},\ }\bibfield  {title} {\enquote
  {\bibinfo {title} {Dynamics of a single free-settling spherical particle
  driven by a laser-induced bubble near a rigid boundary},}\ }\href {\doibase
  10.1103/PhysRevFluids.6.093602} {\bibfield  {journal} {\bibinfo  {journal}
  {Phys. Rev. Fluids}\ }\textbf {\bibinfo {volume} {6}},\ \bibinfo {pages}
  {093602} (\bibinfo {year} {2021})}\BibitemShut {NoStop}%
\bibitem [{\citenamefont {Gonzalez-Avila}, \citenamefont {Denner},\ and\
  \citenamefont {Ohl}(2021)}]{Ohl}%
  \BibitemOpen
  \bibfield  {author} {\bibinfo {author} {\bibfnamefont {S.~R.}\ \bibnamefont
  {Gonzalez-Avila}}, \bibinfo {author} {\bibfnamefont {F.}~\bibnamefont
  {Denner}}, \ and\ \bibinfo {author} {\bibfnamefont {C.-D.}\ \bibnamefont
  {Ohl}},\ }\bibfield  {title} {\enquote {\bibinfo {title} {The acoustic
  pressure generated by the cavitation bubble expansion and collapse near a
  rigid wall},}\ }\href {\doibase 10.1063/5.0043822} {\bibfield  {journal}
  {\bibinfo  {journal} {Phys. Fluids}\ }\textbf {\bibinfo {volume} {33}},\
  \bibinfo {pages} {032118} (\bibinfo {year} {2021})}\BibitemShut {NoStop}%
\bibitem [{\citenamefont {Rattray}(1951)}]{Rattray}%
  \BibitemOpen
  \bibfield  {author} {\bibinfo {author} {\bibfnamefont {M.}~\bibnamefont
  {Rattray}},\ }\emph {\bibinfo {title} {Perturbation Effects in Cavitation
  Bubble Dynamics}},\ \href@noop {} {Ph.D. thesis},\ \bibinfo  {school}
  {California Institute of Technology} (\bibinfo {year} {1951})\BibitemShut
  {NoStop}%
\bibitem [{\citenamefont {Plesset}\ and\ \citenamefont
  {Chapman}(1970)}]{Milton}%
  \BibitemOpen
  \bibfield  {author} {\bibinfo {author} {\bibfnamefont {M.~S.}\ \bibnamefont
  {Plesset}}\ and\ \bibinfo {author} {\bibfnamefont {R.~B.}\ \bibnamefont
  {Chapman}},\ }\bibfield  {title} {\enquote {\bibinfo {title} {Collapse of an
  initially spherical vapour cavity in the neighbourhood of a solid
  boundary},}\ }\href {\doibase 10.1017/S0022112071001058} {\bibfield
  {journal} {\bibinfo  {journal} {J. Fluids Mech.}\ }\textbf {\bibinfo {volume}
  {47}},\ \bibinfo {pages} {283--290} (\bibinfo {year} {1970})}\BibitemShut
  {NoStop}%
\bibitem [{\citenamefont {Brennen}(1995)}]{Brennen}%
  \BibitemOpen
  \bibfield  {author} {\bibinfo {author} {\bibfnamefont {C.~E.}\ \bibnamefont
  {Brennen}},\ }\href@noop {} {\emph {\bibinfo {title} {Cavitation and Bubble
  Dynamics}}}\ (\bibinfo  {publisher} {Oxford University Press},\ \bibinfo
  {year} {1995})\BibitemShut {NoStop}%
\bibitem [{\citenamefont {Obreschkow}\ \emph {et~al.}(2011)\citenamefont
  {Obreschkow}, \citenamefont {Tinguely}, \citenamefont {Dorsaz}, \citenamefont
  {Kobel}, \citenamefont {de~Bosset},\ and\ \citenamefont
  {Farhat}}]{obreschkow2011universal}%
  \BibitemOpen
  \bibfield  {author} {\bibinfo {author} {\bibfnamefont {D.}~\bibnamefont
  {Obreschkow}}, \bibinfo {author} {\bibfnamefont {M.}~\bibnamefont
  {Tinguely}}, \bibinfo {author} {\bibfnamefont {N.}~\bibnamefont {Dorsaz}},
  \bibinfo {author} {\bibfnamefont {P.}~\bibnamefont {Kobel}}, \bibinfo
  {author} {\bibfnamefont {A.}~\bibnamefont {de~Bosset}}, \ and\ \bibinfo
  {author} {\bibfnamefont {M.}~\bibnamefont {Farhat}},\ }\bibfield  {title}
  {\enquote {\bibinfo {title} {Universal scaling law for jets of collapsing
  bubbles},}\ }\href {\doibase 10.1103/PhysRevLett.107.204501} {\bibfield
  {journal} {\bibinfo  {journal} {Phys. Rev. Lett.}\ }\textbf {\bibinfo
  {volume} {107}},\ \bibinfo {pages} {204501} (\bibinfo {year}
  {2011})}\BibitemShut {NoStop}%
\bibitem [{\citenamefont {Ory}\ \emph {et~al.}(2000)\citenamefont {Ory},
  \citenamefont {Yuan}, \citenamefont {Prosperetti}, \citenamefont {Popinet},\
  and\ \citenamefont {Zaleski}}]{ory2000growth}%
  \BibitemOpen
  \bibfield  {author} {\bibinfo {author} {\bibfnamefont {E.}~\bibnamefont
  {Ory}}, \bibinfo {author} {\bibfnamefont {H.}~\bibnamefont {Yuan}}, \bibinfo
  {author} {\bibfnamefont {A.}~\bibnamefont {Prosperetti}}, \bibinfo {author}
  {\bibfnamefont {S.}~\bibnamefont {Popinet}}, \ and\ \bibinfo {author}
  {\bibfnamefont {S.}~\bibnamefont {Zaleski}},\ }\bibfield  {title} {\enquote
  {\bibinfo {title} {Growth and collapse of a vapor bubble in a narrow tube},}\
  }\href {\doibase 10.1063/1.870381} {\bibfield  {journal} {\bibinfo  {journal}
  {Phys. Fluids}\ }\textbf {\bibinfo {volume} {12}},\ \bibinfo {pages}
  {1268--1277} (\bibinfo {year} {2000})}\BibitemShut {NoStop}%
\bibitem [{\citenamefont {Supponen}\ \emph {et~al.}(2016)\citenamefont
  {Supponen}, \citenamefont {Obreschkow}, \citenamefont {Tinguely},
  \citenamefont {Kobel}, \citenamefont {Dorsaz},\ and\ \citenamefont
  {Farhat}}]{Supponen}%
  \BibitemOpen
  \bibfield  {author} {\bibinfo {author} {\bibfnamefont {O.}~\bibnamefont
  {Supponen}}, \bibinfo {author} {\bibfnamefont {D.}~\bibnamefont
  {Obreschkow}}, \bibinfo {author} {\bibfnamefont {M.}~\bibnamefont
  {Tinguely}}, \bibinfo {author} {\bibfnamefont {P.}~\bibnamefont {Kobel}},
  \bibinfo {author} {\bibfnamefont {N.}~\bibnamefont {Dorsaz}}, \ and\ \bibinfo
  {author} {\bibfnamefont {M.}~\bibnamefont {Farhat}},\ }\bibfield  {title}
  {\enquote {\bibinfo {title} {Scaling laws for jets of single cavitation
  bubbles},}\ }\href {\doibase 10.1017/jfm.2016.463} {\bibfield  {journal}
  {\bibinfo  {journal} {J. Fluids Mech.}\ }\textbf {\bibinfo {volume} {802}},\
  \bibinfo {pages} {263--293} (\bibinfo {year} {2016})}\BibitemShut {NoStop}%
\bibitem [{\citenamefont {Toro}(1997)}]{Toro}%
  \BibitemOpen
  \bibfield  {author} {\bibinfo {author} {\bibfnamefont {E.~F.}\ \bibnamefont
  {Toro}},\ }\href@noop {} {\emph {\bibinfo {title} {Riemann Solvers and
  Numerical Methods for Fluid Dynamics}}}\ (\bibinfo  {publisher} {Springer
  Berlin Heidelberg},\ \bibinfo {year} {1997})\BibitemShut {NoStop}%
\bibitem [{\citenamefont {Trummler}\ \emph {et~al.}(2020)\citenamefont
  {Trummler}, \citenamefont {Bryngelson}, \citenamefont {Schmidmayer},
  \citenamefont {Schmidt}, \citenamefont {Colonius},\ and\ \citenamefont
  {Adams}}]{Trummler}%
  \BibitemOpen
  \bibfield  {author} {\bibinfo {author} {\bibfnamefont {T.}~\bibnamefont
  {Trummler}}, \bibinfo {author} {\bibfnamefont {S.~H.}\ \bibnamefont
  {Bryngelson}}, \bibinfo {author} {\bibfnamefont {K.}~\bibnamefont
  {Schmidmayer}}, \bibinfo {author} {\bibfnamefont {S.~J.}\ \bibnamefont
  {Schmidt}}, \bibinfo {author} {\bibfnamefont {T.}~\bibnamefont {Colonius}}, \
  and\ \bibinfo {author} {\bibfnamefont {N.~A.}\ \bibnamefont {Adams}},\
  }\bibfield  {title} {\enquote {\bibinfo {title} {Near-surface dynamics of a
  gas bubble collapsing above a crevice},}\ }\href {\doibase
  10.1017/jfm.2020.432} {\bibfield  {journal} {\bibinfo  {journal} {J. Fluid
  Mech.}\ }\textbf {\bibinfo {volume} {899}},\ \bibinfo {pages} {A16} (\bibinfo
  {year} {2020})}\BibitemShut {NoStop}%
\end{thebibliography}
\end{document}